\documentclass{IEEEtaes}

\usepackage{color,array,amsthm}
\usepackage{graphicx}
\usepackage{amsmath,amssymb,amsfonts}
\usepackage{algorithmic}
\usepackage{graphicx}
\usepackage{textcomp}
\usepackage{bm}
\usepackage{subfigure}
\usepackage{cleveref}
\usepackage[dvipsnames]{xcolor}
\usepackage{tabularx}
\usepackage{multirow}

\colorlet{ali}{black}
\colorlet{jeremy}{black}

\newcommand{\angstrom}{\mbox{\normalfont\AA}}

\jvol{XX}
\jnum{XX}
\jmonth{XXXXX}
\paper{1234567}
\pubyear{2024}
\doiinfo{TAES.2024.Doi Number}

\begin{document}

\title{Neural-based Video Compression on Solar Dynamics  Observatory Images} 

\author{Atefeh Khoshkhahtinat}
\affil{West Virginia University, WV USA} 

\author{Ali Zafari}
\affil{West Virginia University, WV USA}

\author{Piyush M. Mehta}
\affil{West Virginia University, WV USA}

\author{Nasser M. Nasrabadi}
\affil{West Virginia University, WV USA}

\author{Barbara J. Thompson}
\affil{NASA Goddard Space Flight Center, MD USA}

\author{Michael S. F. Kirk}
\affil{NASA Goddard Space Flight Center, MD USA}

\author{Daniel E. da Silva}
\affil{NASA Goddard Space Flight Center, MD USA}

\markboth{KHOSHKHAHTINAT ET AL.}{\uppercase{Neural-based Video Compression on Solar Dynamics  Observatory Images}}
\maketitle

\begin{abstract}
{\color{ali}
NASA's Solar Dynamics Observatory (SDO) mission collects extensive data to monitor the Sun's daily activity. In the realm of space mission design, data compression plays a crucial role in addressing the challenges posed by limited telemetry rates. The primary objective of data compression is to facilitate efficient data management and transmission to work within the constrained bandwidth, thereby ensuring that essential information is captured while optimizing the utilization of available resources. This paper introduces a neural video compression technique that  achieves a high compression ratio for the SDO's image data collection. The proposed approach focuses on leveraging both temporal and spatial redundancies in the data, leading to a more efficient compression. In this work, we introduce an architecture based on the Transformer model, which is specifically designed to capture both local and global information from input images in an effective and efficient manner. Additionally, our network is equipped with an entropy model that can accurately model the probability distribution of the latent representations and improves the speed of the entropy decoding step. The entropy model leverages a channel-dependent approach and utilizes checkerboard-shaped local and global spatial contexts. By combining the Transformer-based video compression network with our entropy model, the proposed compression algorithm demonstrates superior performance over traditional video codecs like H.264 and H.265, as confirmed by our experimental results. 

}
\end{abstract}


\section{INTRODUCTION}

T{\scshape he} Solar Dynamics Observatory (SDO) \cite{pesnell2012solar} mission, launched by NASA in 2010, has played a pivotal role in our understanding of the Sun's activity. Operating from geosynchronous orbit in space, the SDO mission collects a significant volume of data,  approximately 1.4 terabytes, every day \cite{schou2012design}. Due to limiting constraints of onboard data storage and telemetry rate,  data compression has become an indispensable tool in space missions. Several papers 
 \cite{fischer2017jpeg2000,zafari2022attention} have explored the implementation of data compression techniques for the SDO mission. In one such study, Fischer \emph{et al.} \cite{fischer2017jpeg2000} emphasize the necessity of onboard data compression to enhance the efficiency of data storage and transmission for extensive solar databases. They applied the JPEG2000 \cite{taubman2002jpeg2000} algorithm to perform lossy image compression on SDO data at various bitrates. Additionally, Zafari \emph{et al.} \cite{zafari2022attention} have specifically designed a neural-based codec for image compression in the SDO mission.

Deep learning studies in the field of image and video compression have gained significant popularity in recent years \cite{yang2022introduction}. Learned image compression methods have shown remarkable performance in comparison to the traditional codecs such as JPEG \cite{wallace1991jpeg}, JPEG2000 \cite{jpeg2000}, and BPG \cite{bpg}. These methods excel in effectively exploiting spatial redundancies within images. Similarly, for video compression, the principles of learned image compression can be extended by extracting and leveraging the temporal correlations between successive frames of a video. This additional element allows a significant enhancement in compression efficiency \cite{ma2019image}. 

The classical technique employed in both image and video compression is transform coding \cite{goyal2001transformcoding}. Codecs which rely on transform coding partition the process of lossy compression into four distinct stages: linear analysis and synthesis transformations, quantization, and entropy coding. Within the realm of deep learning research, each of these stages can be effectively implemented using deep neural networks. The transformation stages, encompassing analysis and synthesis, often adopts a convolutional-based autoencoder which is capable of learning nonlinear functions (transformations) \cite{leshno1993multilayer}. The nonlinear analysis transformation (encoder) has the remarkable potential to map the data into a lower-dimensional latent space, enabling more efficient compression compared to the linear transforms utilized in traditional codecs \cite{balle2021}. The nonlinear synthesis transformation (decoder) acts as an approximate inverse function of the encoder, mapping the latent space to data space to reconstruct the original input data. The entropy model is implemented by utilizing deep generative models \cite{balle2018a,minnen2018,minnen2020,He_2022_CVPR,he2021checkerboard,khoshkhahtinat2023multi} to estimate the probability distribution of the latent representation. Furthermore, various differentiable quantization methods \cite{agustsson2017, balle2016,yang2020bayesquantize,guo2021quantize} are introduced to facilitate end-to-end training of the entire framework.



\begin{figure*}[tp]
    \centering
    \includegraphics[width=0.83\linewidth]{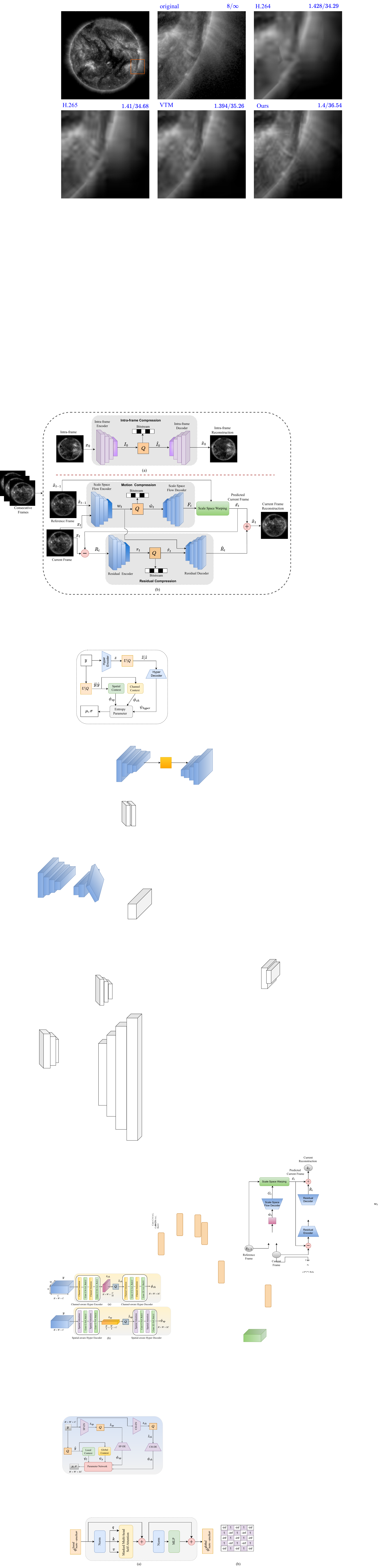}
    \caption{Visual comparison of our proposed neural video compression approach (SSF-HCMWin-ChSp) to other traditional codecs. The performance is reported in terms of bit-rate/distortion [bpp$\downarrow$/PSNR$\uparrow$]. Our proposed method demonstrates lower distortion in terms of PSNR when compared to the other methods, indicating its superior performance in preserving image quality. \emph{Best viewed on screen.}}
    \label{fig:visual-comparison}
\end{figure*}

\textbf{Contributions of This Work}. In this work, we propose a learned video compression approach and apply it to SDO imagery  with the goal of establishing it as the preferred method for on-board data compression in the future space missions. The proposed video compression algorithm effectively leverages both spatial and temporal redundancies inherent in the dataset, which enables the achievement of  high compression ratios. To enhance the capabilities of the non-linear transformation in generating more decorrelated and energy-compacted latents, thereby facilitating a more efficient entropy coding process, we extend the convolutional-based transformation to a Transformer-based architecture. The Transformer block benefits from the self-attention module and locally enhanced block, enabling the capture of both short-range and long-range information. Furthermore, we introduce an entropy model to expedite the decoding process and obtain a more accurate probability distribution of the latent representations. This entropy model is built upon a channel-conditional structure, allowing for the extraction of cross-channel correlations within the latent space. 
In addition, it employs both local and global spatial context models that efficiently and effectively estimate the parameters of the probability distribution for entropy coding. By combining these components, our entropy model significantly contributes to the overall performance of the system, ensuring faster decoding and more precise estimation of the latent representation's probability distribution.

The paper is organized as follows: In Section \ref{sec:related-work}, we present a comprehensive review of learned image and video compression techniques. Additionally, we provide a brief explanation of the entropy modeling component in compression pipeline. In section \ref{sec:methods}, the proposed method is discussed in detail.  Experimental results and ablation studies are provided in section \ref{sec:experiments}. Section \ref{sec:conclusion} concludes the paper.

\section{Related Work}\label{sec:related-work}

\subsection{Neural Image Compression}

The majority of the learning-based image compression methods are  based on the transform coding scheme, which is comprised of four key steps \cite{goyal2001transformcoding}. First, an analysis transform maps the input image to a compact and  decorrelated latent representation. Secondly, a quantization process is performed on the latent representation to obtain a discretized version of the latent representation. In this step, each element of the latent representation vector is rounded to its nearest integer to discard less significant information. Third, entropy coding is employed to generate bitstream of symbols under an entropy model. In the final step, a synthesis transform  maps the quantized latent
representations back to the input image domain \cite{balle2021}.

{Most neural-based image compression networks adopt
the autoencoder architecture \cite{balle2021}, which facilitates the
implementation of an approximately invertible nonlinear
transformation. The encoder component transforms the data (pixels) into a lower-dimensional latent space, while the decoder reverses this process, mapping latents back to pixels. However, while dimensionality reduction is a feature of compression, it's not equivalent. Real compression aims to reduce the entropy of the representation under a shared probability model (the entropy model) between the sender and receiver, rather than just reducing dimensionality. Hence, alongside the transformation network, an entropy model is incorporated to entropy code the latent representation to estimate the bit rate. This process is learned in an end-to-end fashion. However, learning the parameters of these networks poses a challenge due to the non-differentiable nature of quantization, where gradients can either be zero or infinity. To mitigate this concern, several approaches have been presented in the literature to approximate quantization with differentiable functions \cite{agustsson2017, balle2016, yang2020bayesquantize, guo2021quantize}. One common approach is replacing quantization with additive uniform noise \cite{balle2016}, which converts the autoencoder into a variational autoencoder (VAE)~\cite{kingma2013auto} with a uniform encoder. }

{Ball{'{e}} \emph{et al.}\cite{balle2017endtoend} initially introduced the image compression framework as a compressive autoencoder, achieving results comparable to the JPEG2000 \cite{jpeg2000} standard. They integrated the generalized divisive normalization (GDN) function for non-linear transformation and employed a fully-factorized entropy model to estimate the bit-rate of the latent representation. In parallel, other endeavors have aimed to enhance reconstructed image fidelity \cite{agustsson2023multi,mentzer2020high, zhang2018unreasonable}. For instance, Mentzer et al. \cite{mentzer2020high} employed a conditional GAN within an autoencoder-based compression model, elevating perceptual quality. Their proposed framework demonstrated significant advantages for human observers, surpassing BPG even with half the bits.}


\subsection{Entropy Model}

The role of the entropy model is to provide an estimate of the joint probability distribution over the quantized latent representation. The more accurate the entropy model is, the lower the bit-rate in the final compression data stream. Entropy model can enjoy forward and backward adaptation. Forward adaptation utilizes a hyperprior \cite{balle2018a} as a side information to capture spatial dependencies between elements of the latent representation. On the other hand,  backward adaptation employs a context model which leverages the causal neighboring symbols to generate a code for the current symbol. Such a context based method is based on an autoregressive model \cite{minnen2018}  and need a serial processing which slows down the decoding step. In the following section, we
explain both variations in detail.

\paragraph{Forward Adaptation:} The simplest approach to model a distribution over a quantized latent representation $\hat{y}$ is a fully factorized model, \emph{i.e.,} the joint distribution over the latent representation fully factorizes into the product of the marginal distributions. This approach is based on the assumption that all the dimensions of the latent representation are independent from each other:

\begin{align}
\begin{split}
    P_{\bm{\hat{y}}}(\bm{\hat{y}})&=P_{\bm{\hat{y}_1}}(\bm{\hat{y}_1})P_{\bm{\hat{y}_2}}(\bm{\hat{y}_2})\dots P_{\bm{\hat{y}_m}}(\bm{\hat{y}_m})\\
    &=\prod_i P_{\bm{\hat{y}_i}}(\bm{\hat{y}_i}),
\end{split}
\end{align}
where $m$ is the dimension of the latent representation $\bm{\hat{y}}$. The assumption of the full independence, when there are statistical dependencies among the elements of the latent representation, results in suboptimal compression efficiency. Ball{\'{e} \emph{et al.} \cite{balle2018a} adopt a common way to make the entropy model more accurate. They introduce a set of hidden latent variables, denoted as $\hat{z}$, to capture spatial dependencies among elements of the latent representation. It has been proven that multivariate random variables can be independent when they are conditioned on the latent variables \cite{bishop1998latent}. So, if we represent the quantized latent variable as $\hat{z}$, which is also called hyper-prior in the compression literature \cite{balle2018a,balle2021,qian2021globref,qian2022entroformer}, the shared distribution of latent representation between the encoder and decoder can be expressed as follows:
\begin{align}
    P_{\bm{\hat{y}}}(\bm{\hat{y}})
    &=\prod_i P_{\bm{\hat{y}_i}}(\bm{\hat{y}_i}|\bm{\hat{z}}).
    \label{eq:lvm}
\end{align}

\paragraph{Backward Adaptation:} Backward adaptation methods are based on autoregressive generative models \cite{van2016conditional} which define a distribution over the latent representation using the chain rule: 
\begin{align}
\begin{split}
    P_{\bm{\hat{y}}}(\bm{\hat{y}})&=P_{\bm{\hat{y}_1}}(\bm{\hat{y}_1})P_{\bm{\hat{y}_2}}(\bm{\hat{y}_2}|\bm{\hat{y}_{1}})P_{\bm{\hat{y}_3}}(\bm{\hat{y}_3}|\bm{\hat{y}_{1}},\bm{\hat{y}_{2}})\dots\\
    &\dots P_{\bm{\hat{y}_{m-1}}}(\bm{\hat{y}_{m-1}}|\bm{\hat{y}_{<m-1}})P_{\bm{\hat{y}_m}}(\bm{\hat{y}_m}|\bm{\hat{y}_{<m}})\\
    &=\prod_i P_{\bm{\hat{y}_i}}(\bm{\hat{y}_i}|pa(\bm{\hat{y}_i})),
\end{split}
\label{eq:chain-rule}
\end{align}
where $pa(\bm{\hat{y}_i})=\bm{\hat{y}_{<i}}$ shows the parents of $\bm{\hat{y}_i}$, \emph{i.e.}, $\bm{\hat{y}_1},\bm{\hat{y}_2},\dots,\bm{\hat{y}_{i-1}}$. By employing this approach in compression tasks, the distribution of uncoded elements is predicted based on previously decoded elements. It is worth mentioning that in comparison to forward adaptation approaches, backward adaptation methods are bit-free. {This means that backward adaptation methods don't rely on storing additional data to derive distribution parameters for the latent representation. Instead, the required context for each element of the latent representation is acquired from previously decoded elements.}

{The context for computing the conditional probability in backward methods can be classified into two categories: spatial context and channel context. Spatial context models aim to capture correlations between symbols along the spatial axis. Inspired by the PixelCNN \cite{van2016conditional}, initial works \cite{minnen2018} and \cite{lee2018contextadaptive} have utilized a mask convolution layer to exploit causal local latent representations that have already been decoded to estimate the distribution of the current latent element. Using such a local spatial context model introduces the need for serial decoding, which limits parallelization. To enhance the speed of the decoding procedure, He \emph{et al.} \cite{he2021checkerboard} divide the symbols equally into two parts: anchors and non-anchors. The anchors are decoded solely using the hyperprior, while the non-anchors leverage both the hyperprior and context model, a checkerboard convolution is employed to extract context information from the anchor part. This design enables the parallel decoding of both anchor and non-anchor elements.}

{Channel context approaches leverage cross-channel dependencies among latent representation elements. Firstly, Minnen \emph{et al.} \cite{minnen2020} devise a context model to exploit correlations along the channel dimension, which helps to speed up the decoding process. In this approach, the latent representation is split into equal segments along the channel dimension and each segment is predicted using the previously decoded segments. This decoding strategy   leads to a reduction in the number of sequential steps since the decoding time depends on the number of segments. To accelerate the channel-conditional coding, He \emph{et al.} \cite{He_2022_CVPR} adopt an uneven grouping paradigm  that involves organizing the channels into segments with varying sizes. Specifically, the initial segments are assigned fewer channels compared to the later segments. This arrangement is based on the observation that the later channels contain less information compared to the beginning channels.}

\subsection{Neural Video Compression}
The majority of neural video compression algorithms typically employ two types of coding: predictive coding and transform coding \cite{yang2022introduction}.  Predictive coding is utilized in inter-frame predition to reduce temporal redundancy by predicting the current frame based on one or more previously reconstructed frames and then computing the residual or difference between the current frame and the predicted frame. In video compression pipelines, alongside inter-frame coding, intra-frame coding is employed, which focuses on exploiting merely spatial redundancy within a single frame. Intra-frame coding follows similar principles to image compression methods, i.e., transform coding. Thus, to summarize, frames are categorized into three classes  in the video coding literature \cite{rippel2019learned}:

\begin{enumerate}

\item I-frame ("Intra-coded"): {I-frames, also referred to as keyframes, serve as the initial points within a video and undergo independent compression using image compression methods. They don't rely on other frames for both encoding and decoding.}
\item P-frame ("Predicted"): {P-frames belong to the inter-frame group and are predicted from past frames within the video sequence. These past reference frames can be either preceding I-frames or other P-frames used for encoding and decoding. P-frames store only the changes or differences in the currant frame compared to the reference frames, significantly reducing the file size and enhancing transmission efficiency.}
\item B-frames ("Bi-directional"): {B-frames are included in the inter-frame group and are predicted using information from both preceding and subsequent frames. They capture differences between frames before and after them in the sequence, achieving higher compression efficiency compared to P-frames. However, due to their reliance on both preceding and succeeding frames, B-frames become more vulnerable to errors during transmission, affecting their accuracy.} 

\end{enumerate}

In recent years, a variety of neural video codecs have been proposed. Some codecs are specifically designed for low delay settings \cite{lu2019dvc,chen2019learning,agustsson2020scale,lin2020m,liu2020conditional,hu2021fvc,rippel2021elf}, aiming to exploit temporal redundancies by utilizing previous frame or frames to predict the target frame. On the other hand, certain codecs are better suited for random access settings \cite{wu2018video,djelouah2019neural,yang2020learning,pourreza2021extending,yilmaz2021end}, leveraging bidirectional prediction techniques. The low delay setting is particularly well-suited for applications like live streaming, where real-time transmission is crucial. In contrast, the random access setting is more suitable for applications such as playback, where the reference frame can be obtained from the future. In the subsequent paragraphs, we provide an overview of selected works in the field of learned video compression. 

Drawing inspiration from conventional hybrid video compression schemes, Lu \emph{et al.}  \cite{lu2019dvc} proposed the Deep Video Compression (DVC) network, which served as a pioneering end-to-end deep video compression framework.  Their model incorporated the pre-trained Flownet \cite{ranjan2017optical} to estimate optical flow and employed bilinear warping techniques for obtaining motion compensation. For the motion and residual compression, two auto-encoder
based networks are used to compress them. Agustsson \emph{et al.} \cite{agustsson2020scale} introduced Scale-Space Flow (SSF) as a solution to tackle the issue of fast motion in optical flow estimation. In scale-space flow framework, the scale channel is introduced as an uncertainty parameter to Gaussian blur the regions which are susceptible to disocclusions and fast motion. Motivated by the fact that video is essentially a sequence of images characterized by temporal redundancy, several works \cite{habibian2019video,pessoa2020end} have adopted a 3D autoencoder-based framework, i.e., extension of image compression network. This approach aims to eliminate spatial-temporal correlations in video data by using a spatiotemporal transformation.  To improve the performance of such networks,   Habiban \emph{et al.} employ a temporally conditioned entropy model to exploit the temporal dependencies in the latent space.  Hu \emph{et al.} \cite{hu2021fvc} suggested the Feature-space video compression network (FVC) which is an enhanced version of DVC. In this approach, all the crucial components such as motion estimation, motion compression, motion compensation and residual compression are executed in the feature space rather than pixel space. 

\begin{figure*}[tp]
    \centering
    \includegraphics[width=0.85\linewidth]{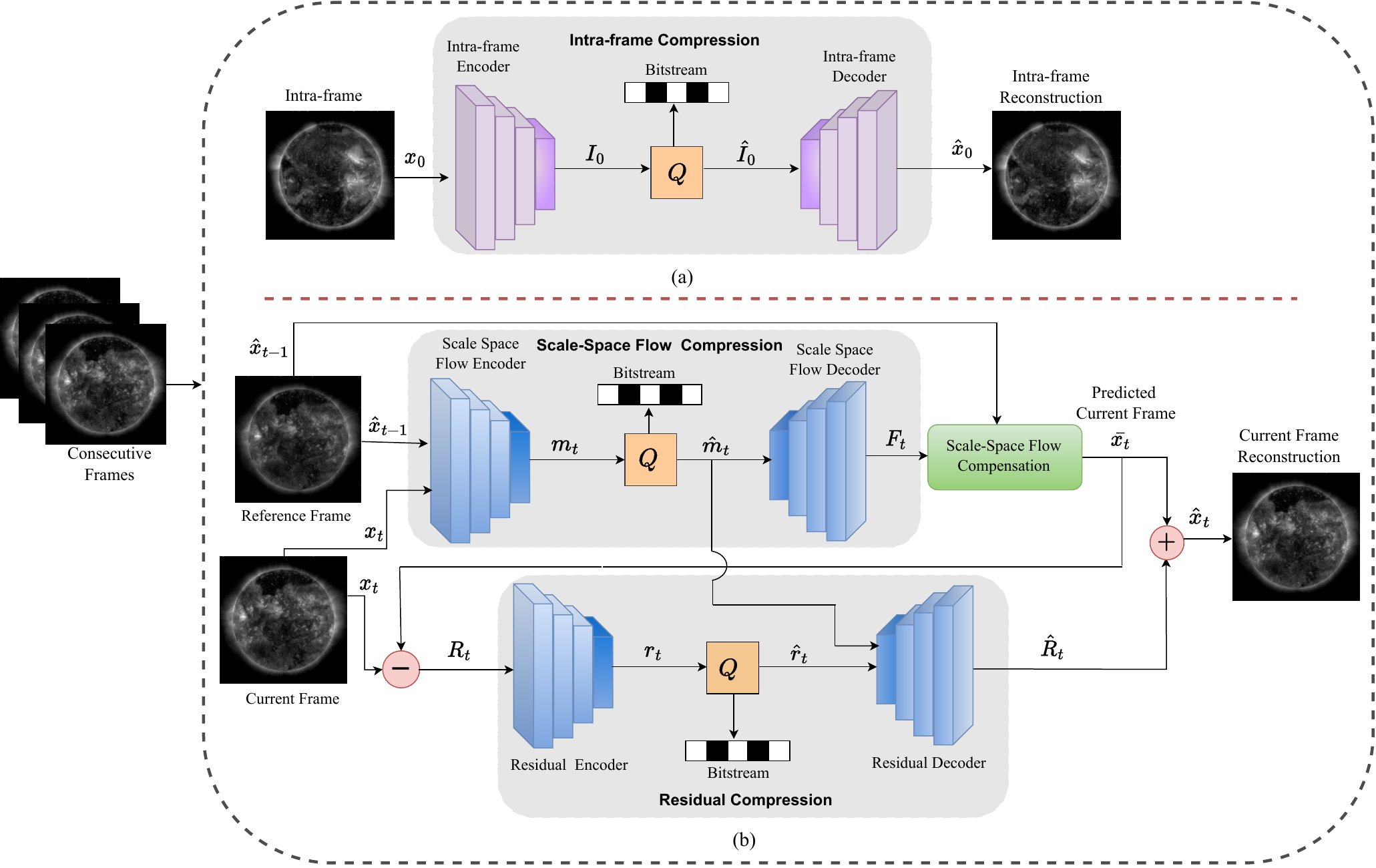}
    \caption{ An overview of our neural video compression network. (a) Architecture of I-frame compression model. (b)  Architecture of P-frame compression model which consists of scale-space flow compression and residual compression networks. First, the motion information and scale field are simultaneously estimated and encoded into a quantized latent representation $\bm{\hat{m}_{t}}$. Second, the previous reconstruction frame $\bm{\hat{x}_{t-1}}$, is warped through the decoded motion and scale fields, $\bm{F_t}$, resulting in the prediction, $\bm{\bar{x}_{t}}$. Third, the residual, $\bm{R_t}$, which is calculated as the difference between the original current frame, $\bm{x_t}$, and the warped prediction, $\bm{\bar{x}_{t}}$, is encoded into a quantized latent representation $\bm{\hat{r}_{t}}$. This residual latent representation is then decoded into $\bm{\hat{R}_{t}}$, which is added to the warped prediction,  to obtain the final reconstructed current frame, $\bm{\hat{x}_{t}}$ = $\bm{\bar{x}_{t}}$ + $\bm{\hat{R}_{t}}$.}
    \label{fig:SSF_Net}
\end{figure*}

While DVC, SSF, and FVC employ a single previous frame as the reference frame, Lin \emph{et al.} \cite{lin2020m} take advantage of multiple previous frames to enhance the accuracy of predicting the current frame. Mentzer \emph{et al.} \cite{mentzer2022neural} developed a neural video compression model based on Generative Adversarial Networks (GANs) \cite{goodfellow2020generative} to improve the perceptual quality of reconstructed frames. Recently, Mentzer \emph{et al.} \cite{mentzer2022vct} introduced a framework that does not explicitly estimate motion. Rather, they used a temporal Transformer for the purpose of entropy modeling, resulting in improved performance.

\subsection{Transformers}

The Transformer \cite{vaswani2017attention} was initially introduced in the field of natural language processing (NLP) and greatly revolutionized this domain. The impressive achievements of the Transformer in NLP have motivated many researchers to adopt the Transformers architecture in a wide range of computer vision (CV) tasks, including object detection \cite{carion2020end,zhu2021dd,zheng2020end}, image classification \cite{touvron2021training,dosovitskiy2021vit}, semantic segmentation \cite{wang2021max,wang2021end,zheng2021rethinking} and many other applications. ViT \cite{dosovitskiy2021vit} is the pioneering study that utilizes stacked Transformer encoders for image classification task and yields outstanding performance when compared to their CNN counterparts. In ViT, an input image is initially  tokenized by dividing it into nonoverlapping patches. Then, these patches are fed into a transformer architecture, where the self-attention mechanisms are employed to capture long-range correlations. Despite the self-attention capability to capture global relationships, it exhibits quadratic growth in computational complexity with the sequence length. Consequently, this results in substantial computational costs and make it infeasible for vision tasks that require high-resolution images. To address this challenge, the Swin Transformer \cite{liu2021swin} is introduced as a solution to reduce the computational complexity of the self-attention mechanism from quadratic to linear by limiting self-attention calculations to within non-overlapping windows. Additionally, the Swin Transformer network is capable of generating hierarchical representations  which are crucial for dense prediction tasks \cite{YuanFHLZCW21}.

\section{Methods}\label{sec:methods}

\subsection{Overview}
Our model builds upon the SSF \cite{agustsson2020scale} network, which is a widely used solution  for low-latency video compression. As illustrated in Figure \ref{fig:SSF_Net}, it consists of two types of compression models: I-frame compression and P-frame compression. The P-frame compression model is further divided into two core components: the scale-space flow compression network and the residual compression network. All three primary networks, namely I-frame compression, scale-space flow compression, and residual compression — rely on the autoencoder architecture \cite{balle2017endtoend}. The SSF network's significant contribution lies in extending optical flow to scale-space flow by introducing a scale field as an additional channel to the motion field. The incorporation of the scale field enables the model to effectively blur regions with disocclusion and fast motion, resulting in enhanced inter-frame prediction. The detailed core components of the P-frame compression model are described as follows: 

\textbf{Scale-Space Flow Compression:} The scale-space flow compression is an autoencoder-based architecture  which employs the current frame $\bm{x_t}$ and previous reconstruction frame $\bm{\hat{{x}}_{t-1}}$ as inputs. The encoder part simultaneously computes and encodes the motion field existing between the two successive frames. At the decoder side, the quantized latent representation of motion is decoded $F_t=(F_x,F_y,F_z)$ and outputs a 2-channel motion field $(F_x,F_y) \in \mathbb{R}^ {2 \times H \times W}$ as well as a one-channel scale field $F_z \in \mathbb{R}^ {H \times W} $.

\textbf{Scale-Space Flow Compensation:} Using the reference frame, which is the previously reconstructed frame $\bm{\hat{{x}}_{t-1}}$, in conjunction with the motion and scale fields, the scale-space flow compensation module aims to predict the current frame $\bm{\bar{x}_t}$. The compensation function is accomplished through the use of a scale-space warping operation. In this process, the reference frame is initially convolved progressively with a Gaussian kernel, resulting in a stack of blurred variants of the reference frame:    
\begin{equation}
   \bm{X}=[\bm{{\hat{{x}}_{t-1}}},\bm{{\hat{{x}}_{t-1}}}*G(s_1),...,\bm{{\hat{{x}}_{t-1}}}*G(s_M)], 
\end{equation}
\noindent where $\bm{X} {\in \mathbb{R}^ {H \times W \times {M+1}}}$ represents scale-space volume which includes the reference frame and $M$ blurred versions of it. $G(s_i)$ denotes  the Gaussian kernel with a scale parameter of $s_i$. For every pixel in coordinate $[x, y]$, the motion-compensated pixel is computed as:
\begin{align}
\begin{split}
   &\bm{\bar{x}_t} =  \text{Scale-Space-Warp}(\bm{\hat{{x}}_{t-1}},\bm{F})\\
   &{\bm{\bar{x}_t}[x,y]}=\bm{X}[x+\bm{F_{x}}[x,y],y+\bm{F_{y}}[x,y],\bm{F_{z}}[x,y]].
\end{split}
\end{align}
Trilinear interpolation is applied in the scale-space volume to predict pixels, which is equivalent to jointly performing bilinear warping and blurring on the reference frame.

\textbf{Residual Compression:} Following the scale-space flow compensation step, the residual between the original frame and its prediction is computed and fed into the residual compression network, which is constructed upon the autoencoder framework \cite{balle2017endtoend}. Similar to any compression approach, the encoder part can eliminate spatial redundancies within the residual and transform it into a decorrelated representation $\bm{r_t}$. Subsequently, the latent representation of the residual is quantized and further entropy encoded into a bit-stream file, utilizing information from its estimated distribution. { On the decoder side, the residual can be reconstructed from the entropy-decoded $\hat{r_t}$ and a latent representation $m_t$.}

\begin{figure*}[tp]
    \centering
    \includegraphics[width=0.85\linewidth]{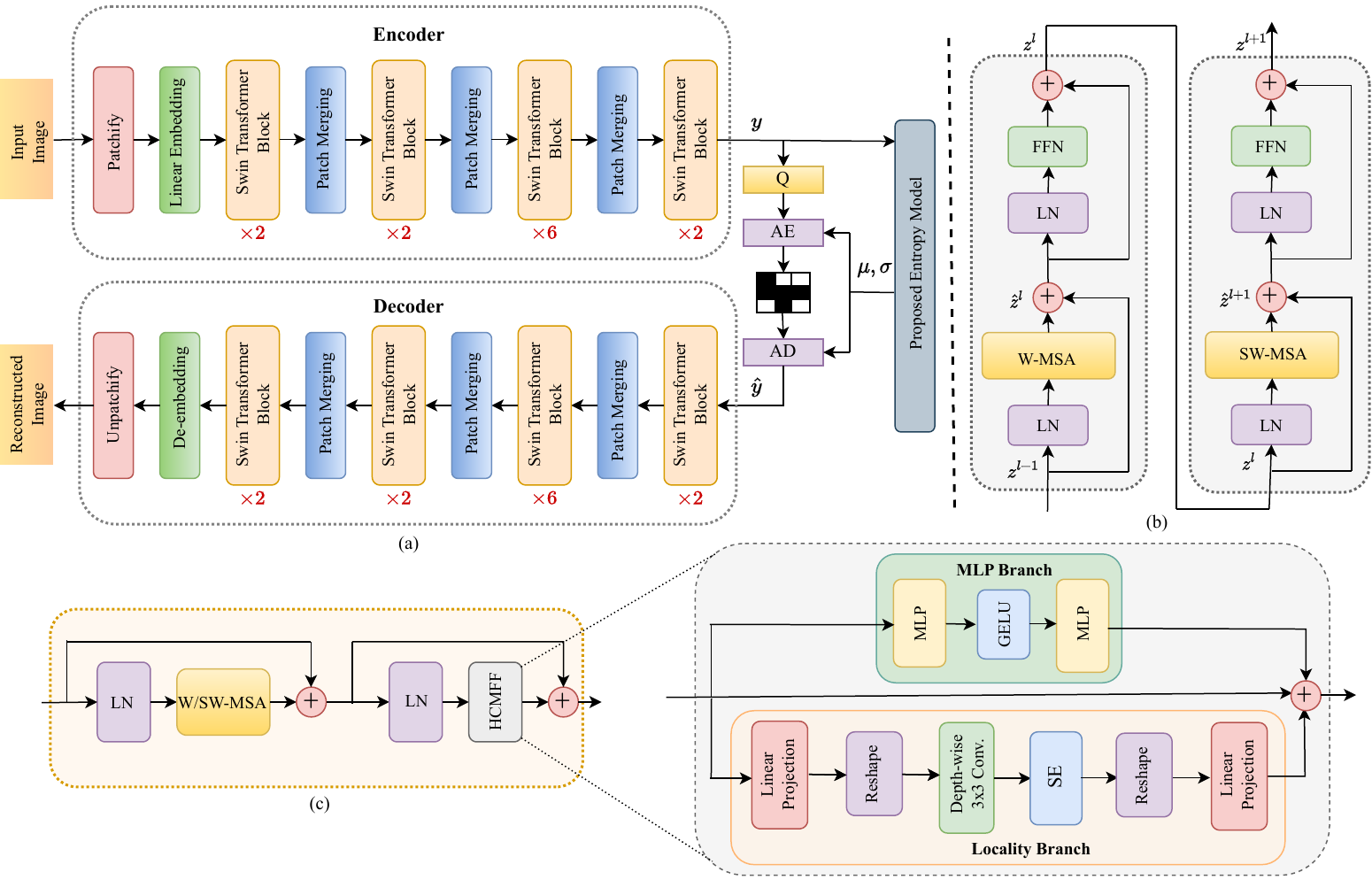}
    \caption{{\color{ali}{(a) Swin Transformer-based architecture designed to compress the I-frame, scale-space flow and residual. Within this framework, Q performs scalar quantization, while AE and AD denote the arithmetic encoder and decoder, respectively. The checkerboard box represents the bitstream file of the compressed image, consisting of zeros and ones. The parameters $\mu$ and $\sigma$ are predicted by the proposed entropy model to estimate the probability distribution of the quantized latent representation $\bm{\hat{y}}$. (b) Two successive Swin Transformer blocks. (c) Hybrid CNN-MLP Window (HCMWin) Transformer block.}}}
    \label{fig:Swin Trans}
\end{figure*}

\subsection{ Transformer-based Autoencoder}
Inspired by the success of Transformer architectures in computer vision, we have utilized the Swin Transformer to construct the encoders and decoders of the Scale-Space Flow network. Figure \ref{fig:Swin Trans}(a) illustrates the architecture based on the Swin Transformer, which is employed for compressing the I-frame, residual, and scale-space flow. Furthermore, we have extended the Swin Transformer block to our proposed Hybrid CNN-MLP Window (HCMWin) Transformer block, which enhances the preservation of local information.

\subsubsection{Swin Transformer-based Encoder}The original Swin Transformer \cite{liu2021swin}, which is
employed as an encoder, consists of four modules: Patchify,
Linear Embedding, Swin Transformer block, and Patch
merging. First, the image ${x\in \mathbb{R}^{ 
C_{in}\times H \times W}}$  is split into
non-overlapping patches through the patchify block to generate a sequence of patches. If the size of each patch is $N \times N$, the total number of patches become $\frac{H}{N} \times \frac{W}{N}$. In the
second step, these patches are flattened and projected into an
embedding space with dimension $C$ by a linear embedding
layer. Then, the output of these two blocks is passed through the
multiple Swin Transformer blocks and patch merging layers. Swin Transformer blocks
preserve the number of patches, \emph{i.e.}  $\frac{H}{N} \times \frac{W}{N}$ and are able
to extract semantic features by computing local self-attention
within each non-overlapping window. The patch
merging layer is responsible to generate hierarchical feature maps by reducing the resolution 
of feature maps and increasing the channels dimension of feature maps. Patch merging groups each
$ 2\times2 $ adjacent patches, which results in reducing the number of patches by
factor of $4$, and concatenates them depth-wise. After, it applies
a linear layer on the resulted patches to obtain
the output feature map with the desired channel number. 
\subsubsection{Swin Transformer-based Decoder }
The Swin Transformer decoder is designed as the inverse symmetric counterpart of the encoder. To achieve this, we make specific replacements in the decoder. These replacements involve replacing the patchify block with an unpatchify block, the patch merging layer with the patch splitting layer, and the linear embedding layer with a de-embedding layer.
\subsubsection{Swin Transformer block}
The Swin Transformer block is the fundamental module of the Swin Transformer architecture. In contrast to the traditional vision Transformer block, a window-based multi-head self-attention (W-MSA) and a shifted-window-based multi-head self-attention (SW-MSA) are utilized to build the Swin Transformer block. W-MSA partitions the input image into non-overlapping windows and then computes the self-attention within local windows which results in reducing the computational complexity from quadratic to the linear with respect to the number of tokens. Despite the significant reduction in computational complexity, W-MSA ignores the relationships between different windows. To address this challenge, shift-window-based multi-headed self-attention (SW-MSA) is placed after the W-MSA module. Therefore, the number of Swin Transformer blocks is always even, where one block uses W-MSA, and the other one employs SW-MSA . Figure \ref{fig:Swin Trans}(b) shows two consecutive Swin Transformer blocks which are made up of layer normalization (LN), window based multi-head self-attention (W-MSA) or shift-window-based multi-headed self-attention (SW-MSA), residual connection and a 2-layer MLP with a Gaussian Error Linear Unit (GELU) as the nonlinear activation function. The process of successive Transformer blocks are formulated as:

\begin{equation}
 \begin{aligned}
    &\bm{\hat z ^\mathnormal{l}} = {W\mbox{-}MSA(LN\bm(\bm{z ^{{\mathnormal{l}}-1}}}))+\bm{z ^{{\mathnormal{l}}-1}},
    \\& \bm{z ^\mathnormal l} = {FFN(LN(\bm{\hat z ^\mathnormal{l}}))}+ \bm{\hat z ^\mathnormal{l}},\\ &  \bm{\hat{z} ^{{\mathnormal{l}}+1}} = {SW\mbox{-}MSA(LN( \bm{z ^\mathnormal l}))}+\bm{z ^\mathnormal l},\\& \bm{{z} ^{{\mathnormal{l}}+1}} = {MLP(LN(\bm{\hat{z} ^{{\mathnormal{l}}+1}}))}+\bm{\hat{z} ^{{\mathnormal{l}}+1}},  
\end{aligned}
\end{equation}

\noindent where $\bm{\hat z ^\mathnormal{l}} $ and $\bm{\hat{z} ^{{\mathnormal{l}}+1}}$ represent  the outputs of W-MSA and SW-MSA blocks, respectively. The used self-attention in W-MAS and SW-MSA can be expressed as follows:
\begin{equation}
 \begin{aligned}
 Attention(\bm{Q},\bm{K}, \bm{V})= softmax( \frac{\bm{Q} \bm{{K}^T}}{\sqrt{d}}+\bm{B}) \bm{V},
 \end{aligned}
\end{equation}

\noindent where $Q$, $K$ and $V\in \mathbb{R}^{M^2 \times d
}$ represent the query, key and value matrices respectively, $d$ is the dimension of key and $M^2$ denotes the number of patches in a window. $B$ shows the learnable  relative position encoding which are obtained from the bias matrix $B' \in \mathbb{R}^{{(2M-1)}\times{(2M-1)}} $ with Learnable parameters. If the number of heads is $K$, the attention mechanism is conducted $K$ times in parallel, and the results for all heads are concatenated. Subsequently, a linear projection is applied to the concatenated outputs to obtain the final result.

\subsubsection{HCMWin Transformer Block}
The  Hybrid CNN-MLP Window (HCMWin) Transformer architecture is built by replacing the traditional feed-forward network in the Swin Transformer block with our proposed Hybrid CNN-MLP feed forward module. This modification allows the extraction of both long-range and short-range dependencies,  which is particularly important for image compression tasks that require capturing local information.

{The introduced Hybrid CNN-MLP Feed-Forward Network (HCMFF) consists of an MLP branch, resembling the conventional feed-forward network in ViT, and a locality branch. By combining window-based self-attention modules with the Hybrid CNN-MLP feed-forward network, the HCMWin Transformer block effectively captures both global and local context information. Window-based self-attention excels at capturing long-range information, while the locality branch within the feed forward network specializes in capturing short-range information. As depicted in Figure \ref{fig:Swin Trans}(c), the output of the second-layer normalization is fed into the HCMFF, which then is directed into the MLP and Locality branches. These two branches are executed in parallel, with their outputs subsequently being summed together.}


\begin{figure}
    
  \centering
  \scalebox{0.58}{\includegraphics{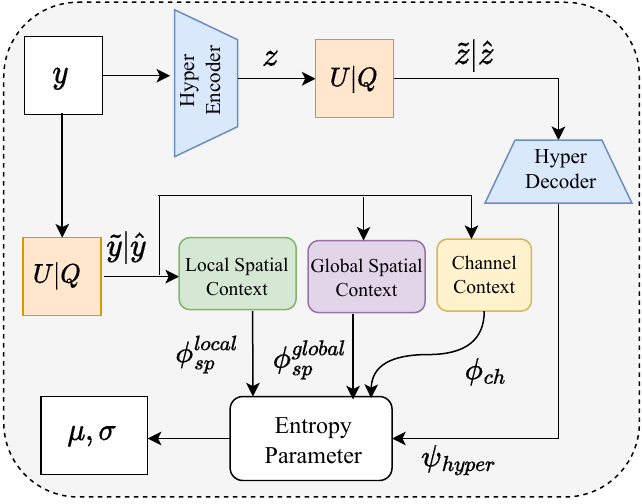}}
  \caption{Diagram of the proposed entropy model. Entropy coding operation for hyper-prior is excluded for simplicity.}
  \label{fig:High Leval Diagram}
\end{figure}

\begin{figure*}[tp]
    \centering
    \includegraphics[width=0.76\linewidth]{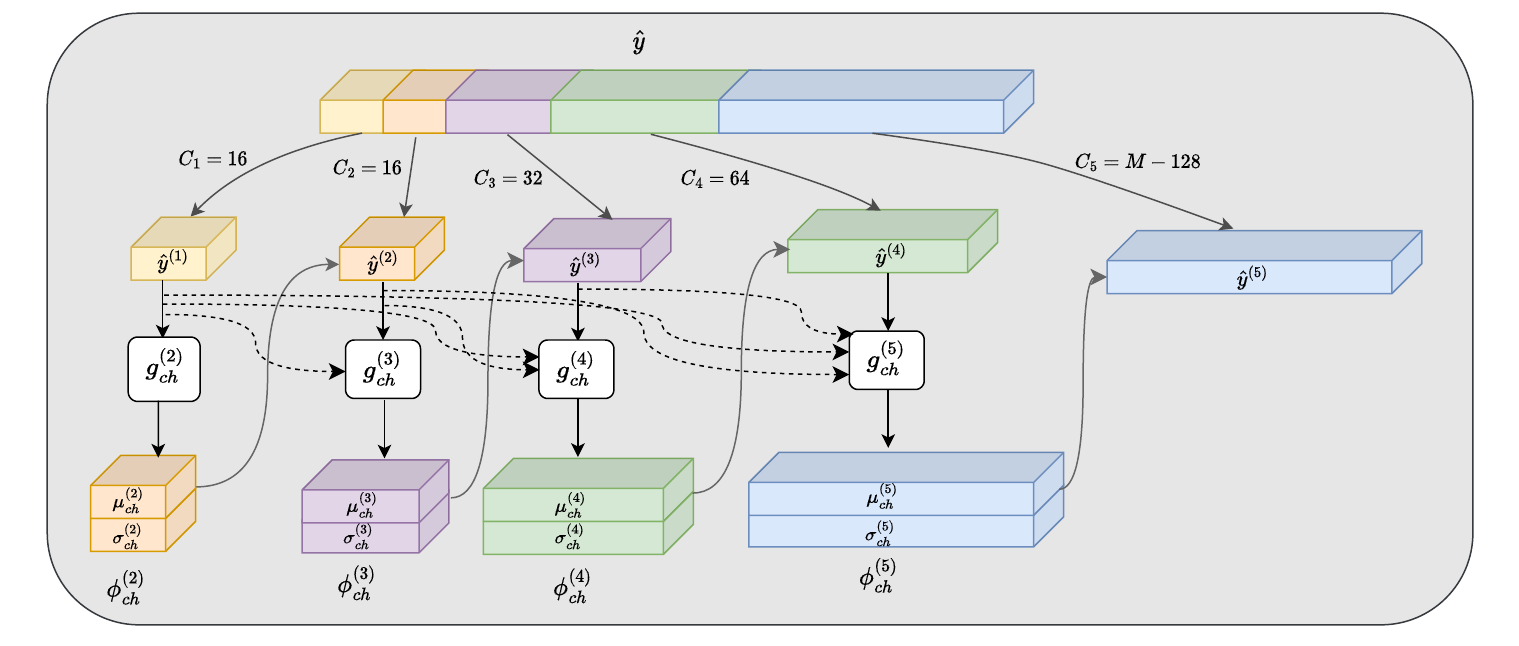}
    \caption{{\color{ali}Channel-conditional coding with uneven
grouping channels. The quantized latent $\bm{\hat{y}}$, which comprises $M$ channels, is segmented into five slices along the channel dimension.  The decoded slice $\bm{\hat{y}_{i}}$ helps the decoding of future slices.}}
    \label{fig:Channel Conditional}
\end{figure*}

{As previously mentioned, the extraction of local information occurs within the locality branch. As shown in Figure \ref{fig:Swin Trans}(c), in the locality branch, each token initially
undergoes a pass through a linear projection layer to expand its dimension. Subsequently, the tokens are reshaped into a 2D token map, which is suitable for subsequent layers. A depth-wise convolution with a kernel size of 3 × 3 is then applied to extract local information and inject inductive bias into the network. Afterward, the SE \cite{hu2018squeeze} module, refers to the Squeeze and Excitation block, is employed to model inter-channel dependencies. Finally, the 2D token maps are flattened  and fed into an additional linear layer to project them onto the input channel dimension.}




\subsection{Proposed Entropy Model}

\begin{figure*}[tp]
    \centering
    \includegraphics[width=0.76\linewidth]{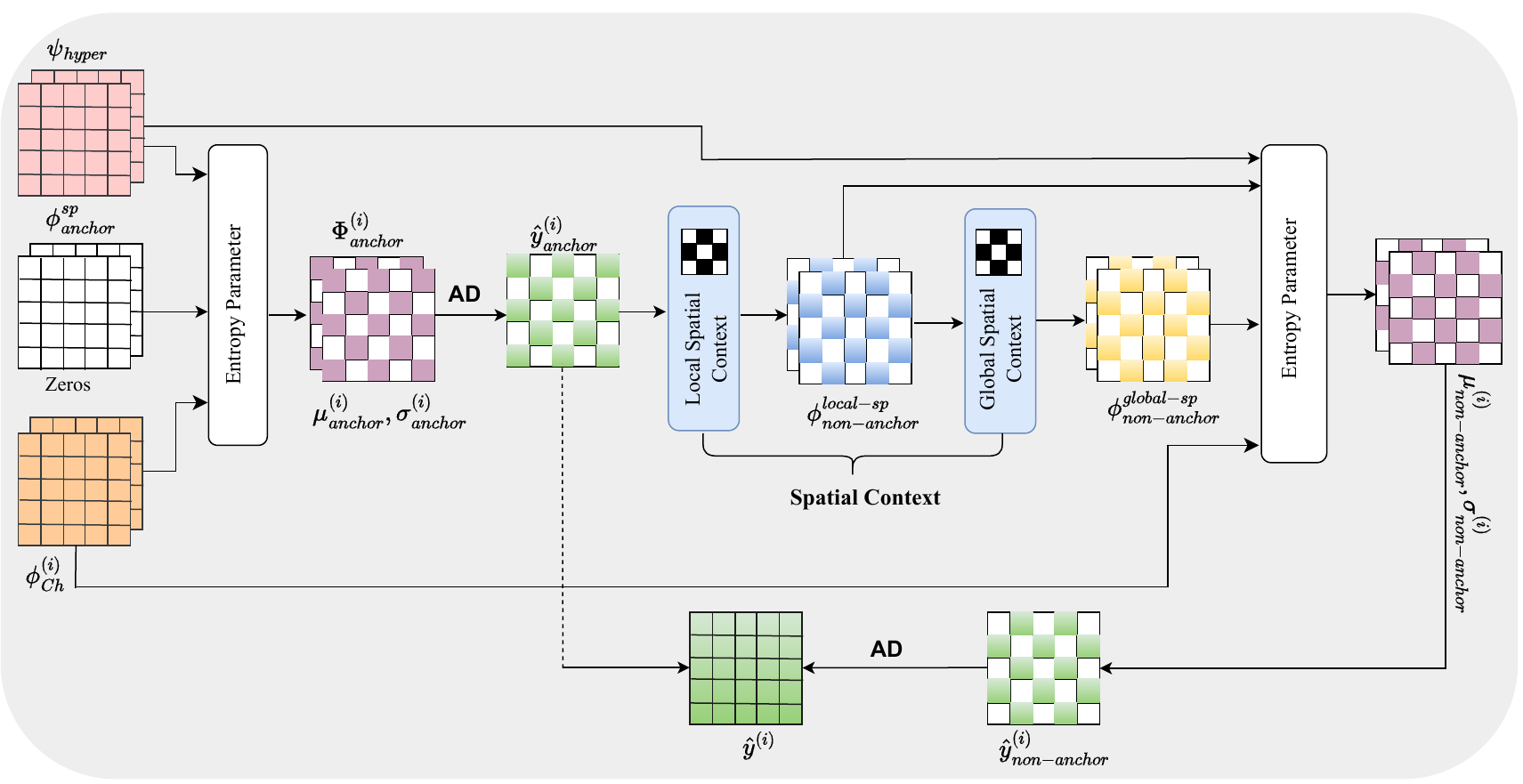}
    \caption{{\color{ali} The implementation of our proposed entropy model within the $i$-th segment. Local spatial context is constructed based on the checkerboard-shaped convolution. Global spatial context is comprised of masked attention module. In the first pass,  the distribution parameters of all the anchor elements are estimated in parallel by using solely the hyperprior and channel context. In the second pass, the non-anchor group leverages the hyperprior, as well as both the spatial and channel contexts, to obtain the entropy parameters. AD denotes arithmetic decoder. }}
    \label{fig:Two-ways Decod}
\end{figure*}

Our proposed entropy model inherits the   advantage of both the hyperprior, which serves as an implementation of forward adaptation, and autoregressive entropy models that employ backward adaptation modeling. Figure \ref{fig:High Leval Diagram} illustrates the high-level structure of our proposed entropy model, comprising the hyper autoencoder, context models (including local spatial, global spatial, and channel blocks), and the parameter network which generates the distribution parameters. Building upon the previous work \cite{minnen2018}, we model the conditional probability of each latent variable as a univariate Gaussian with its mean and standard deviation convolved with a unit uniform distribution:

\begin{equation}
 \begin{aligned}
 &P_{\bm{\hat{y}}|\bm{\hat{z}}}(\bm{\hat{y}}|{\bm{\hat{z}},\bm{\theta}})=\prod_{i=1}(\mathcal{N}(\mu_{i},\,{\sigma_{i}}^2) \ast \mathcal{U}(-\frac{1}{2},\frac{1}{2}))(\hat{y}_i),\\
    & 
    (\mu_i,\sigma_i)=g_{ep}(g_{ls}(\bm{\hat{y}_{<i}^{(k)}}),g_{gs}(\bm{\hat{y}_{<i}^{(k)}}),g_{ch}(\bm{\hat{y}^{<k}}),h_s(\bm{\hat{z}})),
\end{aligned}
\end{equation}

\noindent where mean $\mu_{i}$ and standard deviation  $\sigma_{i}$ are determined by the entropy model. $\bm{\hat z}$ is the quantized hyperprior and $\bm{\theta}$ denotes the parameters of the entropy model. $g_{ep}(.)$, $g_{ls}(.)$, $g_{gs}(.)$ , $g_{ch}(.)$, and $h_s(.)$ correspond to entropy parameter function, local spatial context, global spatial context, channel context, and  hyperprior decoder network, respectively. Additionally, $k$ shows the number of channel segment, and $\bm{\hat{y}^{<k}}=\{{{\bm{\hat{y}^{<1}},..., \bm{\hat{y}^{(k-1)}}}}\}$ represents the previously decoded channel segments.
 
The context of proposed entropy model aims to exploit both cross-channel and spatial correlations, while speeding up the decoding procedure. Following the ELIC framework \cite{He_2022_CVPR}, the latent representation is unevenly partitioned into several segments to leverage channel-wise interactions. The latent $\bm{\hat{y}}$, consisting of $M$ channels, is partitioned along the channel dimension into five groups, namely $16$, $16$, $32$, $64$, and $M-128$ channels, respectively. Segment $\bm{\hat{y}^{(i)}}$  depends on all the previously decoded segments. Figure \ref{fig:Channel Conditional} shows the channel-conditional model with uneven allocation.

Figure \ref{fig:Two-ways Decod} depicts our proposed entropy model applied in the $i$-th segment. In each segment, spatial context is employed to extract both local and global spatial correlations. To enhance the speed of decoding step, we adopt the two-ways parallel context model. To do so, each segment is divided into two parts along the spatial axes. The first part, referred to as the anchor, is decoded using  the hyperprior and channel-wise context and does not benefit from spatial context. The second part, called non-anchor, utilizes the hyperprior and both the spatial and channel contexts to estimate the parameters of the probability distribution.

\begin{figure}[tp]
    \centering
    \includegraphics[width=1.\linewidth]{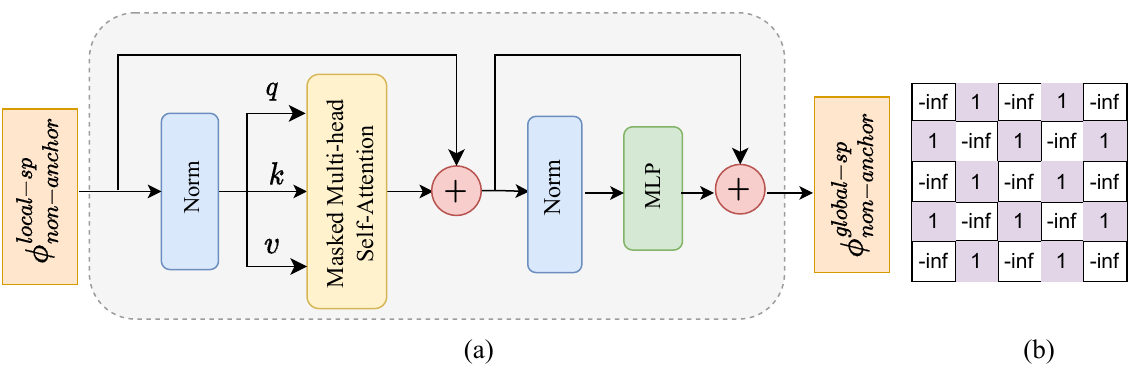}
    \caption{Global Spatial Context Block: (a) The Transformer-based Architecture. (b) An example of checkerboard-shaped mask. 
    Purple squares represent non-anchor positions, while white squares correspond to anchor positions.}
    \label{fig:global context}
\end{figure}

The spatial context component is able to output both local and global contexts. We incorporate the checkerboard pattern within the attention mechanism to capture global correlations and utilize a checkerboard-shaped convolution to leverage local relationships. The use of the checkerboard pattern in the spatial context facilitates parallel decoding, involves two passes. In the initial decoding pass, the distribution parameters of the anchor part are computed in parallel, utilizing only the hyperprior and channel context (its spatial contexts are set to zero). This enables the decoding of the non-anchor part in the subsequent pass, which is also performed in parallel. For decoding the non-anchor part, the neighboring latents of each non-anchor element are convolved with checkerboard shaped kernels to compute the local context.  Then, these computed local contexts of the non-anchor part are then fed into a Transformer-based block, which employs a masked multi-head self-attention module to generate the global context. As shown in Figure \ref{fig:global context}, the normalized local context are considered as query, key and value tokens and passed through the multi-head attention which can be formulated as:\\
\begin{equation}
 \begin{aligned}
    & Attention(\bm{q}, \bm{k}, \bm{v})= Concat(head_1,..., head_m)\bm{W},\\
    & head_i(\bm{q_i},\bm{k_i}, \bm{v_i})= softmax(\bm{q_i} \bm{{k_i}^T}\odot \bm{M})\bm{v_i},
\end{aligned}
\end{equation}

\noindent where $q_i ,k_i, v_i \in \mathbb{R}^{
 HW \times \frac{C}{h} }$ are the queries, keys, and values for the $i$-th head, respectively. $M\in \mathbb{R}^{HW \times HW}$ represents a checkerboard-shaped mask.
 
\subsection{Training Strategy}
\subsubsection{Loss Function}
To train the proposed video compression framework, I-frame and P-frame models are learned jointly with the rate-distortion loss. The total loss over the video clip with sequence length of T can be defined as:
\begin{equation}
D+\lambda R = \sum_{t=0}^{T-1} d({x_t},\hat{{x_t}})+ \lambda [R({I_0})+\sum_{t=1}^{T-1} R({m_t})+ R({r_t})],
\end{equation}

\noindent where $D$ corresponds to the distortion metric, such as Mean Squared Error (MSE) between the original and reconstructed frame. $R$ represents the required bit-rate to encode the quantized latent representation into a binary file. $\bm{I_0}$, $\bm{m_t}$ and $\bm{r_t}$ denotes the latent representations of intra-frame, scale-space flow and residual respectively. $\lambda$ is the Lagrange coefficient which governs the trade-off between rate and distortion. 

The rate term of each latent representation comprises the estimated entropy of the latent representation and the hyperprior, because the probability distribution of each latent representation is conditionally estimated by the hyperprior. Hence, the rate term can be defined as below:
\begin{equation}
 \begin{aligned}
 R(I_0) &= \bm{E}[-\log_2P_{\bm{\hat{I}_{0}}|\bm{\hat{z}_{I}}}(\bm{\hat{I}_{0}}|\bm{\hat{z}_{I}})-\log_2P_{\bm {\hat{z}_{I}}}({\bm{\hat{z}}_{I}})],\\
  R(w_t) &= \bm{E}[-\log_2P_{\bm{\hat{m}_{t}}|\bm{\hat{z}_{m}}}(\bm{\hat{m_t}}|\bm{\hat{z}_{m}})-\log_2P_{\bm {\hat{z}_{m}}}({\bm{\hat{z}}_{m}})],\\
  R(r_t) &= \bm{E}[-\log_2P_{\bm{\hat{r}_{t}}|\bm{\hat{z}_{r}}}(\bm{\hat{r}_t}|\bm{\hat{z}_{r}})-\log_2P_{\bm {\hat{z}_{r}}}({\bm{\hat{z}}_{r}})],\\
\end{aligned}  
\end{equation}

\noindent where $\bm{\hat{z}_{I}}$, $\bm{\hat{z}_{m}}$, and $\bm{\hat{z}_{r}}$ determine the quantized hyper-priors. The non-parametric fully factorized entropy model \cite{balle2017endtoend} is utilized to estimate the probability distribution of hyperpriors.

\subsubsection{Quantization}

In order to enable end-to-end training, the quantization process is replaced with a soft differentiable function during entropy coding. In this study, we add uniform noise to the latent representations, following the approach proposed in \cite{balle2016}, to approximate the hard quantization during the training process. In the test step, a rounding operation, known as hard quantization, is applied to the latent representations.

\section{Experiments}\label{sec:experiments}

\subsection{Dataset} \label{sec:experiments:dataset}
The research conducted in this project is built upon the data collected throughout the SDO mission. The SDO mission gathers images the Sun with three instruments onboard, which operate continuously to capture information. The Helioseismic and Magnetic Imager (HMI) was specifically designed to study oscillations and the magnetic field present in the solar surface, known as the photosphere \cite{schou2012design}. The Atmospheric Imaging Assembly (AIA) captures full-sun images of the solar corona, covering an area of approximately 1.3 solar diameters, with a spatial resolution of around 1 arcsec. It captures images at multiple wavelengths every 12 seconds \cite{lemen2012atmospheric}. In order to gain a deeper understanding of the variations that influence Earth's climate and near-Earth space, the Extreme ultraviolet Variability Experiment (EVE) examines the solar Extreme UltraViolet (EUV) irradiance with high spectral precision \cite{woods2012extreme}. 

Our dataset comprises 138,724 clips, each containing four consecutive images, spanning from 2011 to
2017.

The original SDO dataset has been preprocessed to create a machine learning-ready dataset named SDOML \cite{galvez2019}, which is utilized for this work. AIA images of SDOML dataset are available at wavelengths 94, 131, 171, 193, 211, 304, 335, 1600, 1700  \angstrom\/  with a sampling rate of 6 min. Specifically,  in this paper, AIA images at the wavelength of 94  \angstrom\/  are employed for both the training and testing stages. To train video compression networks using the SDOML dataset, we concatenate four consecutive images to form a temporal chunk consisting of four frames. Our training dataset includes 138,724 clips, each comprising four sequential images, covering the period from 2011 to 2017. During the test phase, we stack 30 consecutive images and create video clips with a Group of Pictures (GoP) size of 30. The rationale for selecting a GoP size of 30 will be elaborated upon in the \textbf{Ablation Study} section.

\subsection{Implementation Details} \label{sec:experiments:implementation}
\subsubsection{Training Settings}
We train our models with the hyper-parameter $\lambda\in\displaystyle\{0.00125, 0.0025, 0.005, 0.01, 0.02, 0.04, 0.08, 0.160, 0.320\}$ to cover a wide range of rate and distortion. The models are trained for $100$ epochs using batches of size $16$. Each batch consists of randomly cropped patches with dimensions of $256 \times 256$, extracted from the original $512 \times 512$ images. The Adam optimizer \cite{kingma15adam} is utilized with an initial learning rate of $10^{-4}$, which is gradually decreased to $1.2\times10^{-6}$ during the training process.

\begin{figure}[tp]
    \centering
    \includegraphics[width=0.93\linewidth]{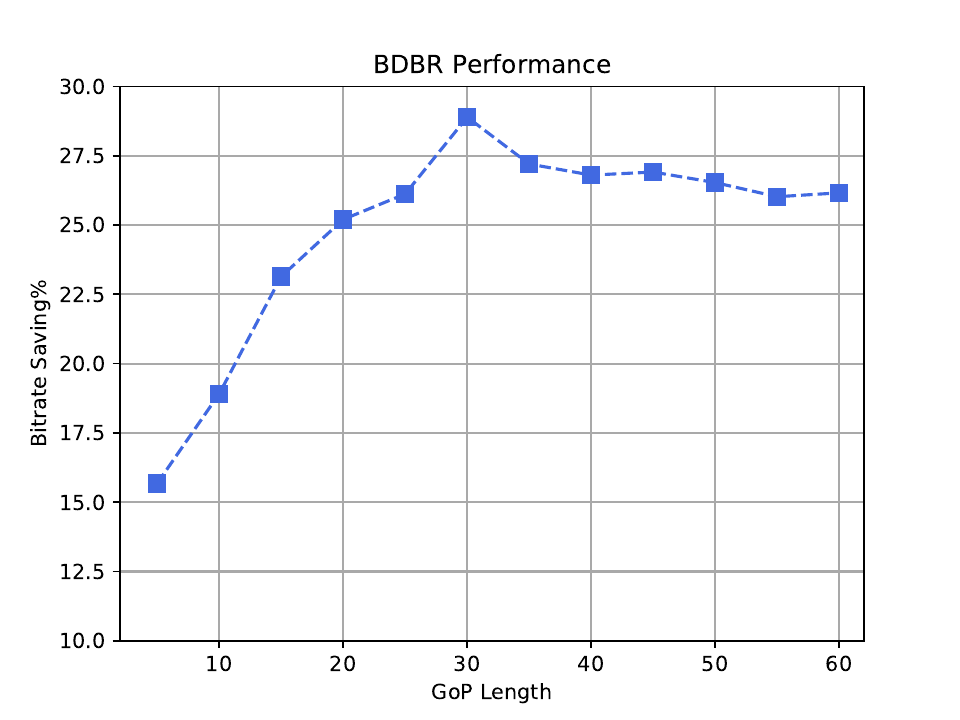}
    \caption{  The bitrate savings achieved by our HCMWin-based video compression compared to neural image compression for different Group of Pictures (GoP) sizes.}
    \label{fig:GOP Test}
\end{figure}

\subsubsection{Traditional Codecs Settings }
To compress data by using H.264 and H.265, we follow the setting in \cite{h264} and use the libx264 and libx265 library, respectively which are provided by the FFmpeg. The command line for applying H.264 and H.265 is as follows:\\
\texttt{{\small ffmpeg -i[input video]-s WxH -c:v libx264| libx265 -crf Q -g GoP -x264|x265-params bframes=0 -preset medium}}\\   
Among them, \emph{W}, \emph{H}, \emph{Q} and \emph{GoP} represent the width of the image, height of the image, quantization parameter, and group of pictures, respectively. \emph{GoP} is set to 30 for our test dataset and \emph{Q} is set as 9, 12, 15, 19, 23, 27, 30 in our settings to obtain compressed files at different bit-rate.
VTM-12.1 software is built from \cite{vtm2022}. The low delay configuration with the highest compression ratio is used. The command line for VTM is as below:\\
\texttt{{\small TAppEncoder -c encoder-lowdelay-main-rext.cfg -InputFile=[input video] --InputBitDepth=8 --OutputBitDepth=8 OutputBitDepthC=8 --DecodingRefreshType=2 -FramesToBeEncoded=[frame number] --SourceWidth=[width] --SourceHeight=[height] --IntraPeriod=32 --QP=[qp] --Level=6.2 }}

\subsection{Ablation Study} 
\subsubsection{ Group of Pictures Size} 

\begin{figure*}[tp]
    \centering
    \includegraphics[width=1\linewidth]{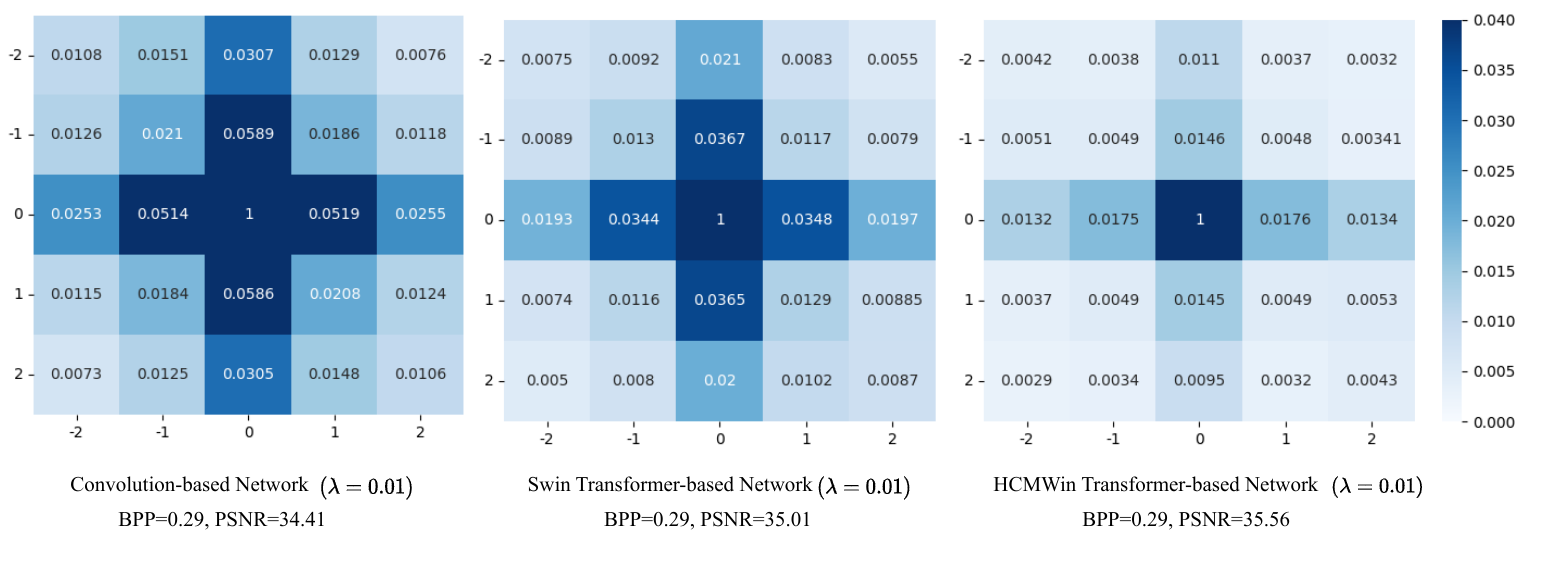}
    \caption{{Comparing the spatial correlation of $\bm{\bar{y}} = \bm{(y - \mu) / \sigma}$ across models trained with $\lambda = 0.01$. HCMWin Transformer-based network outperforms the Swin Transformer-based model in reducing redundancy among latent representation. Both the Swin Transformer and HCMWin Transformer consistently display smaller correlations compared to the Convolution-based Network. To measure the spatial correlation within the latent representation $\bm{y}$, we begin by normalizing the latent vector $\bm{{y}}$ using the estimated mean $\bm{\mu}$ and variance $\bm{\sigma}$ provided by the entropy model. Subsequently, we compute the normalized cross-correlation of $\bm{\bar{y}}$ within a $5\times5$ kernel size in each channel and average these values across all latent channels.} }
    \label{fig:Tran_Ana}
\end{figure*}

\begin{figure}[tp]
    \centering
    \includegraphics[width=0.88\linewidth]{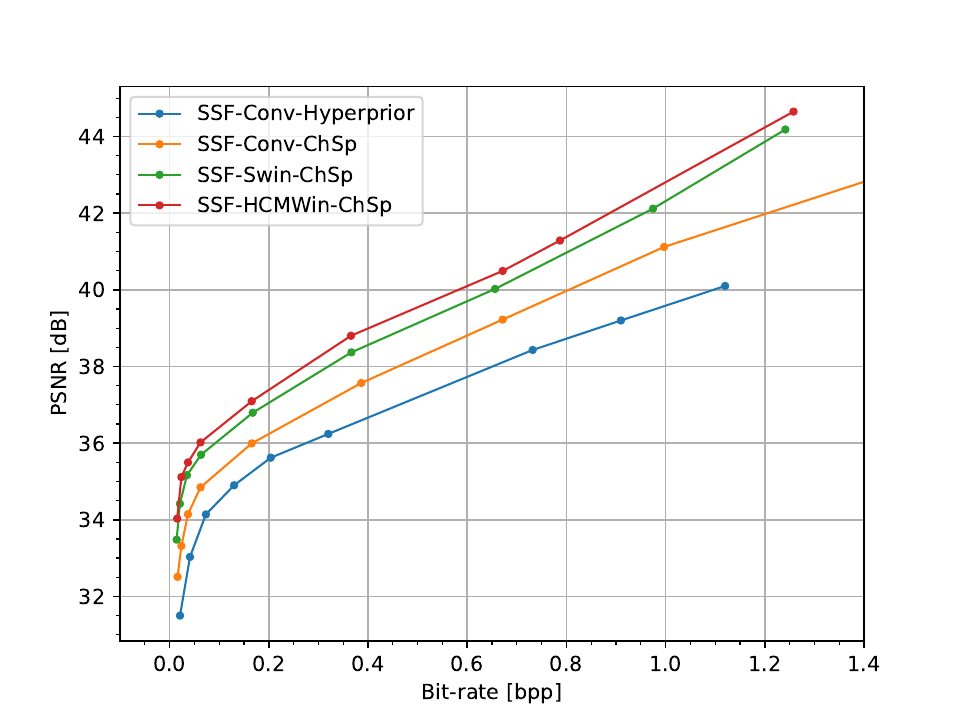}
    \caption{{Comparison of the convolution-based model with two
     different Transformer-based models, namely the Swin Transformer and
    HCMWin Transformer, in terms of the rate-distortion metric. Distortion is measured by PSNR.}}
    \label{fig:Ab-Transformer}
\end{figure}
To determine the optimal GoP size for the testing phase, we experiment with different GoP sizes. We calculate the bit-rate saving of our HCMWin-based video compression method using BDBR (Bitrate Difference to Bitrate Ratio) for each GoP size. This comparison is made against neural image compression, where all frames are treated as intra-frames and compressed using the HCMWin-based I-model.

\begin{figure*}[tp]
    \centering
    \includegraphics[width=\linewidth]{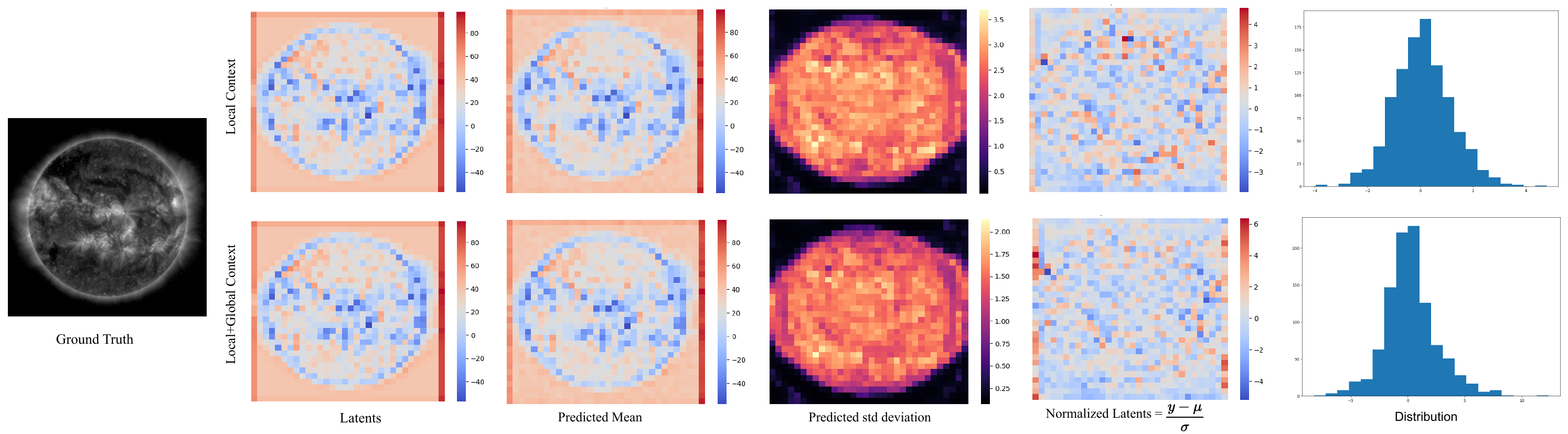}
    \caption{{\color{ali} The figures depict the mechanism of the two variants of the entropy model utilized in the intra-frame model. Each row represents a different model and provides information for the channel with the highest entropy. These figures clearly demonstrate that the integration of both global and local context leads to a  decrease in prediction error and a reduction in the predicted standard deviations. Furthermore, the final normalized latent representation obtained from the ensemble model exhibits a compact distribution, indicating a more accurate entropy model.}}
    \label{fig:ablation_entrop}
\end{figure*}

As depicted in the Figure \ref{fig:GOP Test}, we observe that after reaching the optimal GoP size, the temporal error propagation (this propagation refers to the phenomenon where the reconstruction error of the reference frame propagates to the subsequent frames due to the inter-frame coding approach employed in video compression) becomes significant, and the bit-rate saving doesn't show notable improvements. Based on the experimental results, we set the GoP size for the testing step as 30. We stack 30 consecutive images to create video clips. This choice aims to strike a balance between temporal error propagation and achieving efficient video compression.


\subsubsection{Comparison CNN-based model With Transformer-based models} 

To evaluate the effectiveness of the Swin Transformer architecture and the extension of the Swin Transformer block to the HCMWin Transformer block, we conduct several experiments. We begin with the SSF network as our baseline and  replace its entropy model, consisting of only the hyperprior model, with our proposed entropy model and refer to this model as SSF-Conv-ChSp. Then, we replace the convolutional autoencoders in both the I-frame and P-frame models with the Swin Transformer-based and HCMWin Transformer-based autoencoder and name them SSF-Swin-ChSp and SSF-HCMWin-ChSp, respectively. The results, as depicted in Figure \ref{fig:Ab-Transformer}, reveal several insights. Firstly, integrating our proposed entropy model into SSF significantly improves its efficiency. Additionally, the SSF-Swin-ChSp network surpasses the SSF-Conv-ChSp model in terms of the rate-distortion trade-off. This outcome can be attributed to the inherent limitation of convolutional-based networks to capture long-range correlations. Furthermore, extending the Swin Transformer block to the HCMWin Transformer block helps to further boost the compression performance.

\begin{figure*}[tp]
    \centering
\subfigure{\includegraphics[width=0.48\textwidth]{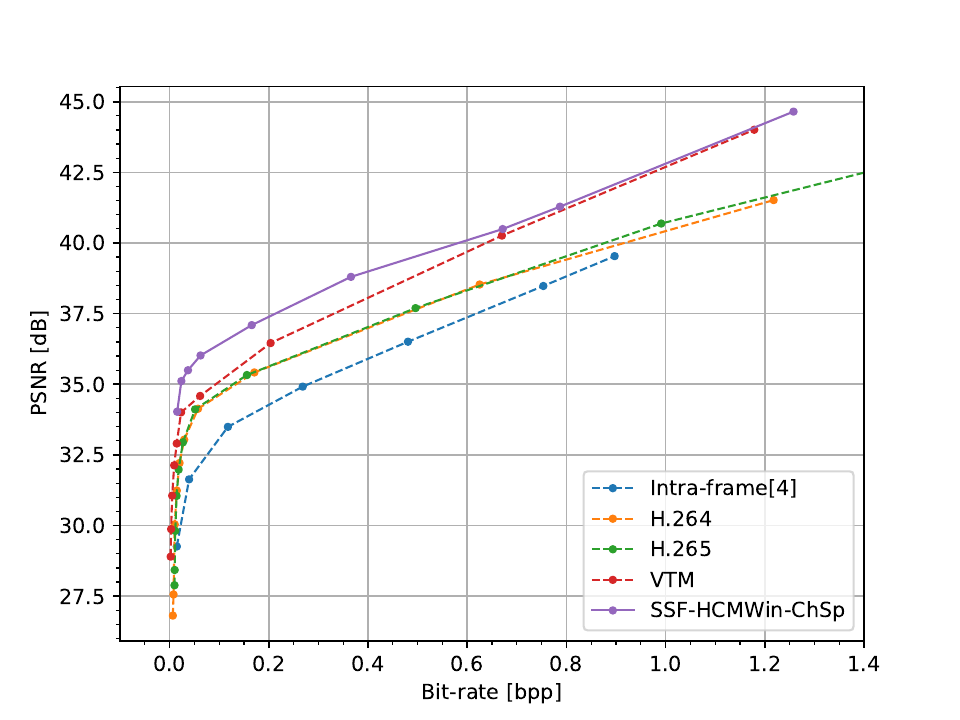}}
    \hfill
    \subfigure{\includegraphics[width=0.47\textwidth]{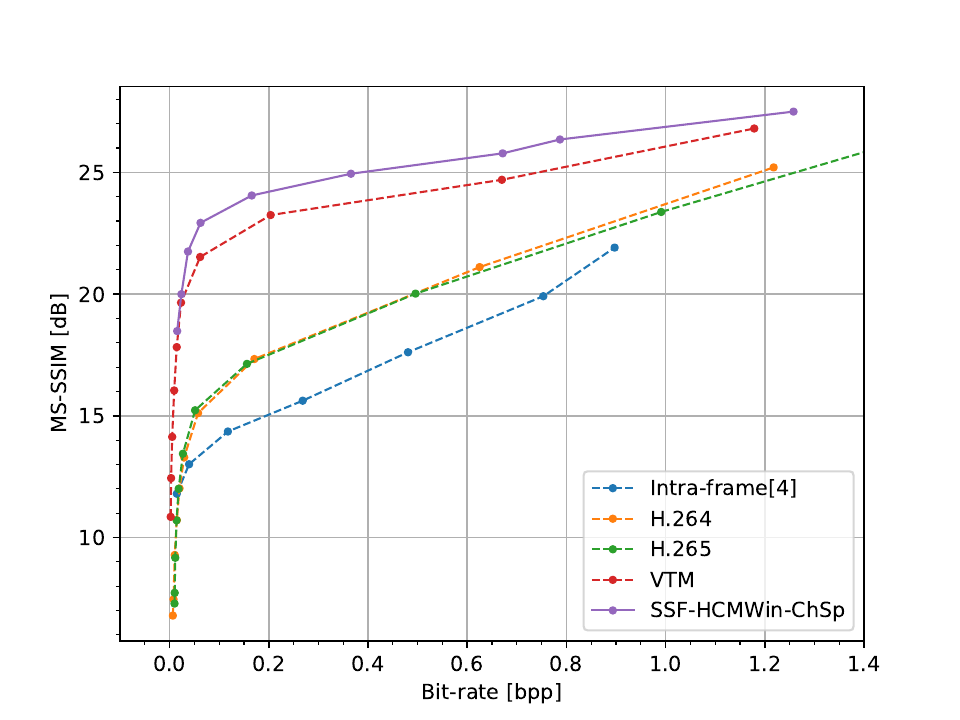}}
    \caption{Rate-distortion curves averaged over the test clips described in Section \ref{sec:experiments:dataset}. On the left, PSNR is calculated from MSE using $10\log_{10}\frac{255^2}{MSE}$. On the right, MS-SSIM is reported in logarithmic scale by $-10\log(1-m)$ to show the differences better, in which $m$ is the MS-SSIM in the range of zero to one.}
    \label{fig:rd-cvrves}
\end{figure*}

}{To explore the efficiency of Transformer architectures in compression from a different angle,  we conducted an analysis of the spatial correlation among latent representations of Intra-frame model in various architectures. As previously stated, the primary objective of the transformation process is to reduce correlations within input data, simplifying quantization and entropy modeling without compromising coding performance. This evaluation involves measuring the correlation among elements in the latent space, particularly focusing on quantifying the correlation between neighboring spatial positions, which tend to exhibit high correlation in natural image sources. Figure \ref{fig:Tran_Ana} illustrates that the Swin and HCMWin Transformeres-based networks surpass Convolutional network in achieving a more decorrelated latent representation. This observation suggests that Transformer-based encoders are more proficient in reducing redundancy across spatial locations compared to convolutional methods. Consequently, they offer a more advantageous trade-off between rate and distortion, enhancing coding efficiency.}

\begin{table}
\caption{ {Comparing bitrate (bpp) and inference latency of entropy parameter estimation during entropy decoding (Dec.) with recent works. All models are trained with $\lambda=0.001$ using a GPU (RTX A6000).([P]:parallel, [S]:Serial)}}
\scalebox{0.87}{
\begin{tabular}{c|cc|c|c}
\hline\hline

\multirow{2}{*} {Method} & \multicolumn{2}{c|}{Context Model}                                                  &  {\multirow{2}{*}{Dec.(ms)} }  &\multirow{2}{*}{bpp}  \\ \cline{2-3} 
                        & Channel            & Spatial &\multicolumn{1}{c|}{}  &                      \\ \hline
Balle \emph{et. al.}\cite{balle2018a}             & -          & -       & $18.4$         & 0.365                        \\                        
Minnen \emph{et. al.}\cite{minnen2018}               & -          & Local[S]        & $>10^3$          & 0.338                        \\
Minnen \emph{et. al.}\cite{minnen2020}                  & Even          & -          & 67          & 0.351                              \\
He \emph{et. al.}\cite{he2021checkerboard}           & -         & Local[P]         & 28.3        & 0.342                \\
Qian \emph{et. al.}\cite{qian2022entroformer}            & -         & Global[S]        &   $>10^3$ &$0.331$                                \\
 He \emph{et. al.}\cite{He_2022_CVPR}         & Uneven         &-          &  37.2      &
$0.358$\\\hline
     
Ours         & Uneven          & Local[S]        &   $>10^3$        & \textbf{0.324}  \\

Ours          & Uneven         & Local[P]          & \textbf{156.9 }         & \textbf{0.335}                  \\

Ours          & Uneven         & Local[P]+Global[P]           & \textbf{162.8}         & \textbf{0.321}                              \\ \hline\hline   

\end{tabular}}

\label{contextmodel}
\end{table}

\subsubsection{Entropy Model}

In this study, we investigate the impact of incorporating checkerboard global context within the uneven grouping channel entropy model. Our findings demonstrate that the inclusion of spatial global context yields superior performance compared to using just local context alone. The evaluation highlights that integrating both global and local context enhances the precision of capturing spatial correlations, leading to a more accurate entropy model. Figure \ref{fig:ablation_entrop} illustrates the efficiency of adding the global spatial context to local spatial context by constructing two different models. The first row represents the model with the local context only, while the second row combines the global context with local context. One image is randomly chosen and encoded using the HCMWin-based Intra-frame model. The highest entropy channels of the two model variants are extracted and visualized in the second column. The predicted means of the entropy model variants are displayed in the third column. It is observed that adding the global context results in a lower prediction error and smaller predicted standard deviations. The last two columns show the normalized latent values, which are obtained by removing the predicted mean and standard deviations, along with their corresponding distributions. The normalized latent of the local+global context entropy model has a more compact distribution, leading to lower bit-rates compared to the local context entropy model.

{Furthermore, we conducted a comparison of the bitrate and inference latency for entropy parameters during entropy decoding of our proposed entropy model compared to other state-of-the-art entropy models.
According to the findings presented in Table \ref{contextmodel}, the speed of our entropy model significantly improves by
employing unevenly grouped channels and implementing a parallel spatial context model, compared to sequential spatial context modeling. Furthermore, our results clearly indicate that incorporating spatial global
context yields superior performance compared to using only local context. The evaluation emphasizes that
integrating both global and local context enhances the precision of capturing spatial correlations, ultimately
resulting in reduced bitrate.}

\begin{figure*}[tp]
    \centering
    \includegraphics[width=1\linewidth]{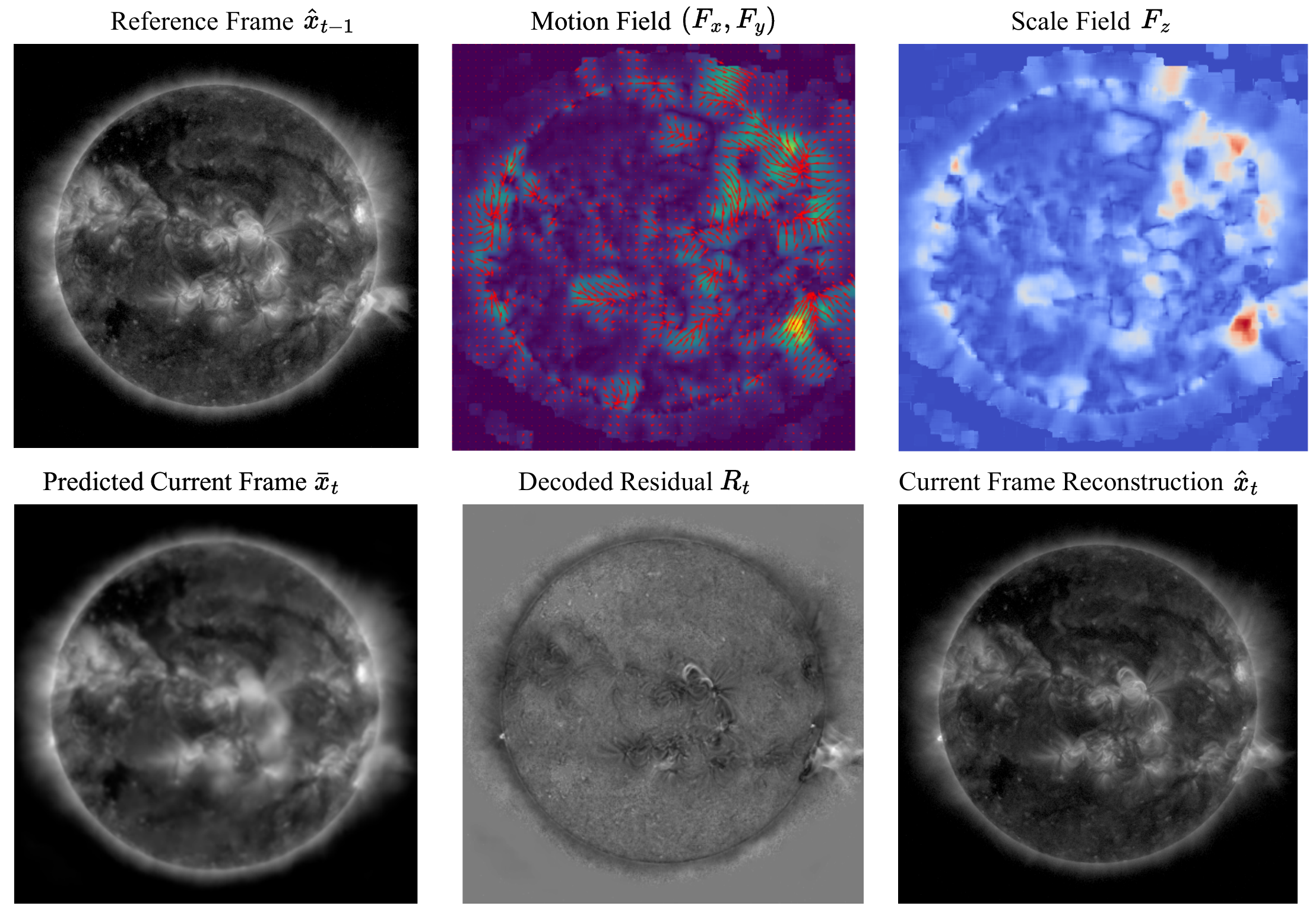}
    \caption{{\color{ali} The visualization of the P-frame model mechanism for two successive images. The scale field demonstrates higher magnitudes in regions where the warping operation may not accurately predict the pixel values, particularly in areas characterized by eruptions. As a result, the network deliberately applies blurring to regions that exhibit complex motion in order to generate reconstructions with simpler residual.  }}
    \label{fig:Mechanism-model}
\end{figure*}

\subsection{Rate-Distortion Performance} 
We compare the rate-distortion (RD) performance of our full model, referred to as SSF-HCMWin-ChSp, with traditional video compression standards and neural image compression on the SDOML dataset \cite{zafari2022attention}, which specifically focuses on intra-frame compression. The distortion are measured by the Peak Signal-to-Noise Ratio (PSNR) metric and Multi-scale structural similarity index Measure (MS-SSIM). By constructing I-frame and P-frame models with a HCMWin Transformer-based autoencoder and employing our proposed entropy model, we achieved a superior RD performance compared to H.264 and H.265, and comparable performance to VTM, as shown in Figure \ref{fig:rd-cvrves}. Our results also highlight that our video compression outperforms image compression on the SDOML dataset, emphasizing the effectiveness of exploiting temporal redundancies. 

{As shown in Figure \ref{fig:rd-cvrves}, the performance of H.265 is comparable to H.264. However, H.265 typically excels in performance at higher resolutions such as 4K or 8K due to its efficient handling of increased detail. Since our dataset operates at a resolution of 512 × 512, which is below the optimal range for H.265, its advantages may not be fully realized. Additionally, our dataset exhibits minimal movement. While H.265's compression techniques are effective in scenarios with substantial motion or intricate changes between frames, these benefits may not be as pronounced in content with minimal movement. Thus, in scenarios like ours with slow or minimal motion, H.265 may not fully leverage its potential compared to situations with dynamic content.}

\begin{table}
\caption{{Comparing the encoding and decoding latencies of our proposed model with other traditional codecs across three distinct bitrate levels (input frame with the size of $512 \times 512$ pixels). Enc and Dec denote encoding and decoding times, respectively, reported in milliseconds.}}
\scalebox{0.77}{
\begin{tabular}{ccclcclcc}
\hline \hline
\multirow{2}{*}{Codec} & \multicolumn{2}{c}{0.13bpp} &  & \multicolumn{2}{c}{0.24bpp} &  & \multicolumn{2}{c}{0.69bpp} \\ \cline{2-3} \cline{5-6} \cline{8-9}
                       & Enc.(ms)          & Dec.(ms)         &  & Enc.(ms)          & Dec.(ms)         &  & Enc.(ms)          & Dec.(ms)         \\ \hline
H.264                 &  33            & 48            &  &39              & 47            &  &  49            & 61           \\
H.265                 & 79             &  87           &  & 95             &  108           &  & 134             & 156            \\
VTM                    & 679              & 763            &  &   753           &  814           &   & 812             & 1004            \\ \hline
\textbf{Ours}                  &              \textbf{891}&  \textbf{1237}           &  &\textbf{972}             & \textbf{1302}            &  &   \textbf{1034}           & \textbf{1513}            \\ \hline \hline
\end{tabular}}
\label{encod_decod}
\end{table} 

\begin{table}
\caption{{Comparing the FLOPs, encoding, and decoding latencies of convolution-based methods and two versions of Transformer-based networks. All models were tested on NVIDIA RTX 6000A using input frame sized at $512\times512$.}}
\scalebox{0.77}{
\begin{tabular}{cclcclccc}
\hline \hline
\multirow{2}{*}{Codec} & \multicolumn{2}{c}{\multirow{2}{*}{Param}} &  & \multicolumn{2}{c}{\multirow{2}{*}{FLOPs}} &  & \multicolumn{2}{c}{Latency} \\ \cline{8-9}
                       & \multicolumn{2}{c}{}                       &  & \multicolumn{2}{c}{}                       &  & Enc.(ms)         & Dec.(ms)         \\ \hline
CNN-based Model                    & \multicolumn{2}{c}{\multirow{3}{*}{}}     96M & & \multicolumn{2}{c}{\multirow{3}{*}{}}     1134.12G &  &   951           &  1291          \\
HCMWin-based Model                & \multicolumn{2}{c}{}   95.84M                 &  & \multicolumn{2}{c}{}  1135.03G                     &  &  997           &  1283                  \\ \hline \hline

\end{tabular}}
\label{complexity}
\end{table}

\subsection{Complexity} 
We evaluated our model's computational complexity against various compression algorithms by conducting experiments to measure encoding and decoding latencies, as detailed in Table \ref{encod_decod}. Traditional handcrafted codecs like H.264, H.265, and VTM were assessed using an Intel Core i7-5930K Broadwell-E CPU-based system. In contrast, our proposed approach underwent testing using a single NVIDIA RTX 6000A graphics card. Additionally, in Table \ref{complexity}, we present the parameters, FLOPs, and encoding and decoding latencies of both convolution-based method and Transformer-based network.It's worth mentioning that all these neural video compression methods consistently utilize our proposed entropy model. The measurements were conducted using input frames with a size of $512\times512$ to ensure uniformity in evaluating these values.

{As reported in Table \ref{encod_decod}, our method achieves better compression performance compared to H.264 and H.265, albeit with an increase in processing time. It is worth mentioning that while the additional time required for the proposed compression method may initially seem prohibitive, its potential benefits in terms of compression quality make it a viable option for space missions where real-time processing speed is not the primary concern. In such missions, images are typically captured over durations longer than 10 seconds \cite{lemen2012atmospheric}, allowing ample time for compression.}

{ Moreover, findings demonstrates that the FLOPs and parameters of the HCMWin Transformer model are nearly equivalent to those of the convolution-based model. This similarity arises from the proposed Transformer's strategy to attain lower complexity through the utilization of window-based attention. This approach enables the Transformer to maintain linear complexity in relation to input resolution, unlike the vanilla ViT Transformer, which exhibits quadratic complexity concerning input resolution.}

\subsection{ {Limitations and Future Work: Onboard Deployment} }

{Considering the constraints of limited power consumption and onboard memory resources, there are potential techniques available to assist in the implementation of Transformer-based models for onboard deployment. These approaches include quantization, network pruning, and knowledge distillation, all aimed at reducing the complexity and memory requirements of the Tranformer-based model while preserving its essential functionality \cite{pan2021scalable}. Quantization of models is a common method used to decrease model size by converting floating-point parameters into lower-bit representations. Several methods, such as MPTQ-ViT \cite{tai2024mptq}, LRP-QViT \cite{ranjan2024lrp}, and PTQ4ViT \cite{yuan2022ptq4vit}, among others, have been proposed for quantizing Transformer models.}

{Network pruning has emerged as a popular approach for eliminating the least important weights in Transformer models, a powerful technique for creating more efficient and compact neural network models without sacrificing performance. Recently, several pruning algorithms have been proposed for Vision Transformers (ViTs), such as UP-ViTs \cite{yu2023unified}, WDPruning \cite{yu2022width}, Evo-ViT \cite{xu2022evo}, and X-Pruner \cite{yu2023x}. Knowledge distillation provides an effective means to address the computing power problem associated with Transformer models by compressing the model size, reducing computational requirements, and enabling efficient deployment in various applications and environments. Knowledge distillation is a promising approach for transferring information from a high-performance teacher to a compact student while maintaining strong performance. The student model, being smaller in size compared to the teacher, requires fewer parameters and less memory, leading to reduced computational requirements during both training and inference. These studies \cite{chen2022dearkd,hao2022learning} propose knowledge distillation approaches applied to Transformer models.}

{In addition, employing hardware accelerators offers numerous benefits for deep learning applications such as Transformer models. The advantages of hardware accelerators include their ability to accelerate computationally expensive operations and their efficiency in managing memory, which are the dominant energy consumers in Transformer models \cite{bhowmick2023optimizing}. GPUs \cite{yu2023boost}, TPUs \cite{reidy2023efficient}, and FPGAs \cite{li2020ftrans,wang2022via} are different types of hardware accelerators used for Transformer models. Our team is actively working on developing technologies for the challenge of onboard deployment which we expect to publish in the near future.}

\subsection{Visualization:}
 In Figure \ref{fig:Mechanism-model}, the mechanism of our P-frame model is illustrated for two consecutive frames. It can be observed that the scale field exhibits larger values in regions where the warping operation may not provide accurate predictions, such as areas with eruptions. Consequently, the network intentionally blurs the regions with complex motion to generate precise reconstructions, leading to sparser residuals.


\section{{Conclusion}}\label{sec:conclusion}

In this work, we have introduced a novel neural video compression framework specifically designed for NASA's SDO mission. Our framework is designed using a Transformer architecture. Additionally, we propose a novel entropy model that not only accelerates the decoding process but also improves the accuracy of the probability distribution estimation for the latent representation. Our experimental results demonstrated a significant enhancement in the compression ratio when applying our video compression approach to the SDO dataset. This improvement can be attributed to the strong temporal correlation observed between the images within the dataset. Moreover, by incorporating the HCMWin-based Transformer and our new entropy model, which leverages channel-wise and both local and global spatial relationships of the latent representation, we have further boosted the compression performance in terms of rate-distortion.


\bibliography{IEEEtaesBIB}

\begin{thebibliography}{10}
\providecommand{\url}[1]{#1}
\csname url@samestyle\endcsname
\renewcommand{\newblock}{\par}
\providecommand{\bibinfo}[2]{#2}
\providecommand{\BIBentrySTDinterwordspacing}{\spaceskip=0pt\relax}
\providecommand{\BIBentryALTinterwordstretchfactor}{4}
\providecommand{\BIBentryALTinterwordspacing}{\spaceskip=\fontdimen2\font plus
\BIBentryALTinterwordstretchfactor\fontdimen3\font minus \fontdimen4\font\relax}
\providecommand{\BIBforeignlanguage}[2]{{%
\expandafter\ifx\csname l@#1\endcsname\relax
\typeout{** WARNING: IEEEtran.bst: No hyphenation pattern has been}%
\typeout{** loaded for the language `#1'. Using the pattern for}%
\typeout{** the default language instead.}%
\else
\language=\csname l@#1\endcsname
\fi
#2}}
\providecommand{\BIBdecl}{\relax}
\BIBdecl

\bibitem{pesnell2012solar}
W.~D. Pesnell, B.~J. Thompson, and P.~Chamberlin
\newblock \emph{The {Solar Dynamics Observatory} {(SDO)}}. Springer, 2012.

\bibitem{schou2012design}
J.~Schou \emph{et~al.}
\newblock  Design and ground calibration of the {Helioseismic} and {Magnetic Imager (HMI)} instrument on the {Solar Dynamics Observatory (SDO)} \newblock  \emph{Solar Physics}, vol. 275, pp. 229--259, 2012.

\bibitem{fischer2017jpeg2000}
C.~E. Fischer, D.~M{\"u}ller, and I.~De~Moortel
\newblock  {JPEG2000} image compression on solar {EUV} images \newblock  \emph{Solar Physics}, vol. 292, pp. 1--22, 2017.

\bibitem{zafari2022attention}
{A. Zafari, A. Khoshkhahtinat, P. M. Mehta, N. M. Nasrabadi, B. J. Thompson, D. Da Silva, and M. S. F. Kirk}
\newblock  Attention-based generative neural image compression on {Solar Dynamics Observatory} \newblock  In \emph{2022 21st IEEE International Conference on Machine Learning and Applications (ICMLA)}. IEEE, 2022, pp. 198--205.

\bibitem{taubman2002jpeg2000}
D.~S. Taubman, M.~W. Marcellin, and M.~Rabbani
\newblock  {JPEG2000}: Image compression fundamentals, standards and practice \newblock  \emph{Journal of Electronic Imaging}, vol.~11, no.~2, pp. 286--287, 2002.

\bibitem{yang2022introduction}
Y.~Yang, S.~Mandt, and L.~Theis
\newblock  An introduction to neural data compression \newblock  \emph{CoRR}, vol. abs/2202.06533, 2022.

\bibitem{wallace1991jpeg}
G.~K. Wallace
\newblock  The {JPEG} still picture compression standard \newblock  \emph{Communications of the ACM}, vol.~34, no.~4, pp. 30--44, 1991.

\bibitem{jpeg2000}
D.~S. Taubman and M.~W. Marcellin
\newblock \emph{{JPEG2000} - image compression fundamentals, standards and practice}, ser. The Kluwer international series in engineering and computer science. Kluwer, 2002.

\bibitem{bpg}
F.~Bellard
\newblock  Bpg image format \newblock  \url{https://bellard.org/bpg/}.

\bibitem{ma2019image}
S.~Ma, X.~Zhang, C.~Jia, Z.~Zhao, S.~Wang, and S.~Wang
\newblock  Image and video compression with neural networks: A review \newblock  \emph{IEEE Transactions on Circuits and Systems for Video Technology}, vol.~30, no.~6, pp. 1683--1698, 2019.

\bibitem{goyal2001transformcoding}
V.~Goyal
\newblock  Theoretical foundations of transform coding \newblock  \emph{IEEE Signal Processing Magazine}, vol.~18, no.~5, pp. 9--21, 2001.

\bibitem{leshno1993multilayer}
M.~Leshno, V.~Y. Lin, A.~Pinkus, and S.~Schocken
\newblock  Multilayer feedforward networks with a nonpolynomial activation function can approximate any function \newblock  \emph{Neural networks}, vol.~6, no.~6, pp. 861--867, 1993.

\bibitem{balle2021}
{J. Ball{\'{e}}, P. A. Chou, D. Minnen, S. Singh, N. Johnston, E. Agustsson, S. J. Hwang, and G. Toderici}
\newblock  Nonlinear transform coding \newblock  \emph{{IEEE} J. Sel. Top. Signal Process.}, 2021.

\bibitem{balle2018a}
J.~Ball{\'{e}}, D.~Minnen, S.~Singh, S.~J. Hwang, and N.~Johnston
\newblock  Variational image compression with a scale hyperprior \newblock  In \emph{6th International Conference on Learning Representations}, 2018.

\bibitem{minnen2018}
D.~Minnen, J.~Ball{\'{e}}, and G.~Toderici
\newblock  Joint autoregressive and hierarchical priors for learned image compression \newblock  In \emph{Advances in Neural Information Processing}, 2018.

\bibitem{minnen2020}
D.~Minnen and S.~Singh
\newblock  Channel-wise autoregressive entropy models for learned image compression \newblock  In \emph{{IEEE} International Conference on Image Processing, {ICIP} 2020, Abu Dhabi, United Arab Emirates, October 25-28, 2020}. {IEEE}, 2020, pp. 3339--3343.

\bibitem{He_2022_CVPR}
D.~He, Z.~Yang, W.~Peng, R.~Ma, H.~Qin, and Y.~Wang
\newblock  {ELIC}: Efficient learned image compression with unevenly grouped space-channel contextual adaptive coding \newblock  In \emph{Proceedings of the IEEE/CVF Conference on Computer Vision and Pattern Recognition (CVPR)}, June 2022, pp. 5718--5727.

\bibitem{he2021checkerboard}
D.~He, Y.~Zheng, B.~Sun, Y.~Wang, and H.~Qin
\newblock  Checkerboard context model for efficient learned image compression \newblock  In \emph{Proceedings of the IEEE/CVF Conference on Computer Vision and Pattern Recognition}, 2021, pp. 14\,771--14\,780.

\bibitem{khoshkhahtinat2023multi}
A.~Khoshkhahtinat, A.~Zafari, P.~M. Mehta, M.~Akyash, H.~Kashiani, and N.~M. Nasrabadi
\newblock  Multi-context dual hyper-prior neural image compression \newblock  \emph{arXiv preprint arXiv:2309.10799}, 2023.

\bibitem{agustsson2017}
{E. Agustsson, F. Mentzer, M. Tschannen, L. Cavigelli, R. Timofte, L. Benini, and L. Van Gool}
\newblock  Soft-to-hard vector quantization for end-to-end learning compressible representations \newblock  In \emph{Advances in Neural Information Processing}, 2017.

\bibitem{balle2016}
J.~Ball{\'{e}}, V.~Laparra, and E.~P. Simoncelli
\newblock  End-to-end optimization of nonlinear transform codes for perceptual quality \newblock  In \emph{2016 Picture Coding Symposium, {PCS} 2016}. {IEEE}, 2016.

\bibitem{yang2020bayesquantize}
Y.~Yang, R.~Bamler, and S.~Mandt
\newblock  Variational bayesian quantization \newblock  In \emph{International Conference on Machine Learning}. PMLR, 2020.

\bibitem{guo2021quantize}
Z.~Guo, Z.~Zhang, R.~Feng, and Z.~Chen
\newblock  Soft then hard: Rethinking the quantization in neural image compression \newblock  In \emph{International Conference on Machine Learning}. PMLR, 2021.

\bibitem{kingma2013auto}
D.~P. Kingma and M.~Welling
\newblock  Auto-encoding variational bayes \newblock  \emph{arXiv preprint arXiv:1312.6114}, 2013.

\bibitem{balle2017endtoend}
J.~Ball{\'{e}}, V.~Laparra, and E.~P. Simoncelli
\newblock  End-to-end optimized image compression \newblock  In \emph{5th International Conference on Learning Representations}. OpenReview.net, 2017.

\bibitem{agustsson2023multi}
E.~Agustsson, D.~Minnen, G.~Toderici, and F.~Mentzer
\newblock  Multi-realism image compression with a conditional generator \newblock  In \emph{Proceedings of the IEEE/CVF Conference on Computer Vision and Pattern Recognition}, 2023, pp. 22\,324--22\,333.

\bibitem{mentzer2020high}
F.~Mentzer, G.~D. Toderici, M.~Tschannen, and E.~Agustsson
\newblock  High-fidelity generative image compression \newblock  \emph{Advances in Neural Information Processing Systems}, vol.~33, pp. 11\,913--11\,924, 2020.

\bibitem{zhang2018unreasonable}
R.~Zhang, P.~Isola, A.~A. Efros, E.~Shechtman, and O.~Wang
\newblock  The unreasonable effectiveness of deep features as a perceptual metric \newblock  In \emph{Proceedings of the IEEE conference on computer vision and pattern recognition}, 2018, pp. 586--595.

\bibitem{bishop1998latent}
C.~M. Bishop
\newblock  Latent variable models. \emph{Learning in graphical models}, vol. 371, 1999.

\bibitem{qian2021globref}
{Y. Qian, Z. Tan, X. Sun, M.Lin, D. Li, Z. Sun, H. Li, and R. Jin}
\newblock  Learning accurate entropy model with global reference for image compression \newblock  In \emph{International Conference on Learning Representations}, 2021.

\bibitem{qian2022entroformer}
Y.~Qian, X.~Sun, M.~Lin, Z.~Tan, and R.~Jin
\newblock  Entroformer: A transformer-based entropy model for learned image compression \newblock  In \emph{International Conference on Learning Representations}, 2022.

\bibitem{van2016conditional}
A.~Van~den Oord, N.~Kalchbrenner, L.~Espeholt, O.~Vinyals, A.~Graves \emph{et~al.}
\newblock  Conditional image generation with {PixelCNN} decoders \newblock  \emph{Advances in neural information processing systems}, vol.~29, 2016.

\bibitem{lee2018contextadaptive}
J.~Lee, S.~Cho, and S.-K. Beack
\newblock  Context-adaptive entropy model for end-to-end optimized image compression \newblock  In \emph{International Conference on Learning Representations}, 2019.

\bibitem{rippel2019learned}
O.~Rippel, S.~Nair, C.~Lew, S.~Branson, A.~G. Anderson, and L.~Bourdev
\newblock  Learned video compression \newblock  In \emph{Proceedings of the IEEE/CVF International Conference on Computer Vision}, 2019, pp. 3454--3463.

\bibitem{lu2019dvc}
G.~Lu, W.~Ouyang, D.~Xu, X.~Zhang, C.~Cai, and Z.~Gao
\newblock  {DVC}: An end-to-end deep video compression framework \newblock  In \emph{Proceedings of the IEEE/CVF Conference on Computer Vision and Pattern Recognition}, 2019, pp. 11\,006--11\,015.

\bibitem{chen2019learning}
Z.~Chen, T.~He, X.~Jin, and F.~Wu
\newblock  Learning for video compression \newblock  \emph{IEEE Transactions on Circuits and Systems for Video Technology}, vol.~30, no.~2, pp. 566--576, 2019.

\bibitem{agustsson2020scale}
E.~Agustsson, D.~Minnen, N.~Johnston, J.~Balle, S.~J. Hwang, and G.~Toderici
\newblock  Scale-space flow for end-to-end optimized video compression \newblock  In \emph{Proceedings of the IEEE/CVF Conference on Computer Vision and Pattern Recognition}, 2020, pp. 8503--8512.

\bibitem{lin2020m}
J.~Lin, D.~Liu, H.~Li, and F.~Wu
\newblock  {M-LVC}: Multiple frames prediction for learned video compression \newblock  In \emph{Proceedings of the IEEE/CVF Conference on Computer Vision and Pattern Recognition}, 2020, pp. 3546--3554.

\bibitem{liu2020conditional}
{J. Liu, S. Wang, W. Ma, M. Shah, R. Hu, P. Dhawan, and R. Urtasun}
\newblock  Conditional entropy coding for efficient video compression \newblock  In \emph{Computer Vision--ECCV 2020: 16th European Conference, Glasgow, UK, August 23--28, 2020, Proceedings, Part XVII}. Springer, 2020, pp. 453--468.

\bibitem{hu2021fvc}
Z.~Hu, G.~Lu, and D.~Xu
\newblock  {FVC}: A new framework towards deep video compression in feature space \newblock  In \emph{Proceedings of the IEEE/CVF Conference on Computer Vision and Pattern Recognition}, 2021, pp. 1502--1511.

\bibitem{rippel2021elf}
O.~Rippel, A.~G. Anderson, K.~Tatwawadi, S.~Nair, C.~Lytle, and L.~Bourdev
\newblock  {ELF-VC}: Efficient learned flexible-rate video coding \newblock  In \emph{Proceedings of the IEEE/CVF International Conference on Computer Vision}, 2021, pp. 14\,479--14\,488.

\bibitem{wu2018video}
C.-Y. Wu, N.~Singhal, and P.~Krahenbuhl
\newblock  Video compression through image interpolation \newblock  In \emph{Proceedings of the European conference on computer vision (ECCV)}, 2018, pp. 416--431.

\bibitem{djelouah2019neural}
A.~Djelouah, J.~Campos, S.~Schaub-Meyer, and C.~Schroers
\newblock  Neural inter-frame compression for video coding \newblock  In \emph{Proceedings of the IEEE/CVF international conference on computer vision}, 2019, pp. 6421--6429.

\bibitem{yang2020learning}
R.~Yang, F.~Mentzer, L.~V. Gool, and R.~Timofte
\newblock  Learning for video compression with hierarchical quality and recurrent enhancement \newblock  In \emph{Proceedings of the IEEE/CVF Conference on Computer Vision and Pattern Recognition}, 2020, pp. 6628--6637.

\bibitem{pourreza2021extending}
R.~Pourreza and T.~Cohen
\newblock  Extending neural {P-frame} codecs for {B-frame} coding \newblock  In \emph{Proceedings of the IEEE/CVF International Conference on Computer Vision}, 2021, pp. 6680--6689.

\bibitem{yilmaz2021end}
M.~A. Y{\i}lmaz and A.~M. Tekalp
\newblock  End-to-end rate-distortion optimized learned hierarchical bi-directional video compression \newblock  \emph{IEEE Transactions on Image Processing}, vol.~31, pp. 974--983, 2021.

\bibitem{ranjan2017optical}
A.~Ranjan and M.~J. Black
\newblock  Optical flow estimation using a spatial pyramid network \newblock  In \emph{Proceedings of the IEEE conference on computer vision and pattern recognition}, 2017, pp. 4161--4170.

\bibitem{habibian2019video}
A.~Habibian, T.~v. Rozendaal, J.~M. Tomczak, and T.~S. Cohen
\newblock  Video compression with rate-distortion autoencoders \newblock  In \emph{Proceedings of the IEEE/CVF International Conference on Computer Vision}, 2019, pp. 7033--7042.

\bibitem{pessoa2020end}
J.~Pessoa, H.~Aidos, P.~Tom{\'a}s, and M.~A. Figueiredo
\newblock  End-to-end learning of video compression using spatio-temporal autoencoders \newblock  In \emph{2020 IEEE Workshop on Signal Processing Systems (SiPS)}. IEEE, 2020, pp. 1--6.

\bibitem{mentzer2022neural}
F.~Mentzer, E.~Agustsson, J.~Ball{\'e}, D.~Minnen, N.~Johnston, and G.~Toderici
\newblock  Neural video compression using gans for detail synthesis and propagation \newblock  In \emph{Computer Vision--ECCV 2022: 17th European Conference, Tel Aviv, Israel, October 23--27, 2022, Proceedings, Part XXVI}. Springer, 2022, pp. 562--578.

\bibitem{goodfellow2020generative}
I.~Goodfellow \emph{et~al.}
\newblock  Generative adversarial networks \newblock  \emph{Communications of the ACM}, vol.~63, no.~11, pp. 139--144, 2020.

\bibitem{mentzer2022vct}
{F. Mentzer, G. Toderici, D. Minnen, S. J. Hwang, S.Caelles, M. Lucic, and E. Agustsson}
\newblock  {VCT}: A video compression transformer \newblock  \emph{arXiv preprint arXiv:2206.07307}, 2022.

\bibitem{vaswani2017attention}
{A. Vaswani, N. Shazeer, N. Parmar, J. Uszkoreit, L. Jones, A. N. Gomez, L. Kaiser, and I. Polosukhin}
\newblock  Attention is all you need \newblock  \emph{Advances in neural information processing systems}, vol.~30, 2017.

\bibitem{carion2020end}
N.~Carion, F.~Massa, G.~Synnaeve, N.~Usunier, A.~Kirillov, and S.~Zagoruyko
\newblock  End-to-end object detection with transformers \newblock  In \emph{Computer Vision--ECCV 2020: 16th European Conference, Glasgow, UK, August 23--28, 2020, Proceedings, Part I 16}. Springer, 2020, pp. 213--229.

\bibitem{zhu2021dd}
X.~Zhu, W.~Su, L.~Lu, B.~Li, X.~Wang, and J.~Dai
\newblock  Deformable transformers for end-to-end object detection \newblock  In \emph{Proceedings of the 9th International Conference on Learning Representations, Virtual Event, Austria}, 2021, pp. 3--7.

\bibitem{zheng2020end}
{M. Zheng, P. Gao, R. Zhang, K. Li, X. Wang, H. Li, and H. Dong}
\newblock  End-to-end object detection with adaptive clustering transformer \newblock  \emph{arXiv preprint arXiv:2011.09315}, 2020.

\bibitem{touvron2021training}
H.~Touvron, M.~Cord, M.~Douze, F.~Massa, A.~Sablayrolles, and H.~J{\'e}gou
\newblock  Training data-efficient image transformers \& distillation through attention \newblock  In \emph{International conference on machine learning}. PMLR, 2021, pp. 10\,347--10\,357.

\bibitem{dosovitskiy2021vit}
{A. Dosovitskiy, L. Beyer, A. Kolesnikov, D. Weissenborn, X. Zhai, T. Unterthiner, M. Dehghani, M. Minderer, G. Heigold, S. Gelly, J.Uszkoreit, and N. Houlsby}
\newblock  An image is worth 16x16 words: Transformers for image recognition at scale \newblock  In \emph{9th International Conference on Learning Representations}. OpenReview.net, 2021.

\bibitem{wang2021max}
H.~Wang, Y.~Zhu, H.~Adam, A.~Yuille, and L.-C. Chen
\newblock  End-to-end panoptic segmentation with mask transformers \newblock  In \emph{Proceedings of the IEEE/CVF conference on computer vision and pattern recognition}, 2021, pp. 5463--5474.

\bibitem{wang2021end}
{Y. Wang, Z. Xu, X. Wang, C.Shen, B. Cheng, H. Shen, and H. Xia}
\newblock  End-to-end video instance segmentation with transformers \newblock  In \emph{Proceedings of the IEEE/CVF conference on computer vision and pattern recognition}, 2021, pp. 8741--8750.

\bibitem{zheng2021rethinking}
{S. Zheng, J. Lu, H. Zhao, X. Zhu, Z. Luo, Y. Wang, Y. Fu, J. Feng, T. Xiang, P. H. S Torr, and L. Zhang}
\newblock  Rethinking semantic segmentation from a sequence-to-sequence perspective with transformers \newblock  In \emph{Proceedings of the IEEE/CVF conference on computer vision and pattern recognition}, 2021, pp. 6881--6890.

\bibitem{liu2021swin}
{Z. Liu, Y. Lin, Y. Cao, H. Hu, Y. Wei, Z. Zhang, S. Lin, and B. Guo}
\newblock  {Swin Transformer}: Hierarchical vision transformer using shifted windows \newblock  In \emph{Proceedings of the IEEE/CVF international conference on computer vision}, 2021, pp. 10\,012--10\,022.

\bibitem{YuanFHLZCW21}
{Y. Yuan, R. Fu, L. Huang, W. Lin, C. Zhang, X. Chen, and J. Wang}
\newblock  {HRFormer}: High-resolution transformer for dense prediction \newblock  \emph{NeurIPS}, 2021.

\bibitem{hu2018squeeze}
J.~Hu, L.~Shen, and G.~Sun
\newblock  Squeeze-and-excitation networks \newblock  In \emph{Proceedings of the IEEE conference on computer vision and pattern recognition}, 2018.

\bibitem{lemen2012atmospheric}
J.~R. Lemen \emph{et~al.}
\newblock  The atmospheric imaging assembly {(AIA)} on the {Solar Dynamics Observatory (SDO)} \newblock  \emph{Solar Physics}, vol. 275, pp. 17--40, 2012.

\bibitem{woods2012extreme}
{T. N. Woods, F. G. Eparvier, R. Hock, A. R. Jones, D. Woodraska, D. Judge, L. Didkovsky, J. Lean, J. Mariska, H. Warren, D. McMullin, P. Chamberlin, G. Berthiaume, S. Bailey, T. Fuller-Rowell, J. Sojka, W. K. Tobiska, and R. Viereck}
\newblock  Extreme ultraviolet variability experiment {(EVE)} on the {Solar Dynamics Observatory (SDO)}: Overview of science objectives, instrument design, data products, and model developments \newblock  \emph{The solar dynamics observatory}, pp. 115--143, 2012.

\bibitem{galvez2019}
{R. Galvez, D. F. Fouhey, M. Jin, A. Szenicer, A. Mu{\~{n}}oz-Jaramillo, M. C. M. Cheung, P. J. Wright, M. G. Bobra, Y. Liu, J. Mason, and R. Thomas}
\newblock  A machine-learning data set prepared from the {NASA} solar dynamics observatory mission \newblock  \emph{The Astrophysical Journal Supplement Series}, may 2019.

\bibitem{kingma15adam}
D.~P. Kingma and J.~Ba
\newblock  Adam: {A} method for stochastic optimization \newblock  In \emph{3rd International Conference on Learning Representations}, 2015.

\bibitem{h264}

\newblock  The {H.264}/{AVC} reference software \newblock  Available at \url{http://iphome. hhi.de/suehring/}, 2022.

\bibitem{vtm2022}

\newblock  Versatile video coding reference software \newblock  Available at \url{https://vcgit.hhi.fraunhofer.de/jvet/VVCSoftware_VTM}, 2022.

\bibitem{pan2021scalable}
Z.~Pan, B.~Zhuang, J.~Liu, H.~He, and J.~Cai
\newblock  Scalable {Vision Transformers} with hierarchical pooling \newblock  In \emph{Proceedings of the IEEE/cvf international conference on computer vision}, 2021, pp. 377--386.

\bibitem{tai2024mptq}
Y.-S. Tai \emph{et~al.}
\newblock  {MPTQ-ViT}: {Mixed-Precision Post-Training Quantization for Vision Transformer} \newblock  \emph{arXiv preprint arXiv:2401.14895}, 2024.

\bibitem{ranjan2024lrp}
N.~Ranjan and A.~Savakis
\newblock  {LRP-QViT: Mixed-Precision Vision Transformer Quantization via Layer-wise Relevance Propagation} \newblock  \emph{arXiv preprint arXiv:2401.11243}, 2024.

\bibitem{yuan2022ptq4vit}
 {PTQ4ViT}: Post-training quantization for vision transformers with twin uniform quantization \newblock

\bibitem{yu2023unified}
H.~Yu and J.~Wu
\newblock  {A unified pruning framework for Vision Transformers} \newblock  \emph{Science China Information Sciences}, vol.~66, no.~7, p. 179101, 2023.

\bibitem{yu2022width}
F.~Yu, K.~Huang, M.~Wang, Y.~Cheng, W.~Chu, and L.~Cui
\newblock  {Width \& depth pruning for Vision Transformers} \newblock  In \emph{Proceedings of the AAAI Conference on Artificial Intelligence}, vol.~36, no.~3, 2022, pp. 3143--3151.

\bibitem{xu2022evo}
 {Evo-ViT}: Slow-fast token evolution for dynamic vision transformer \newblock

\bibitem{yu2023x}
L.~Yu and W.~Xiang
\newblock  {X-pruner: explainable pruning for Vision Transformers} \newblock  In \emph{Proceedings of the IEEE/CVF Conference on Computer Vision and Pattern Recognition}, 2023, pp. 24\,355--24\,363.

\bibitem{chen2022dearkd}
X.~Chen, Q.~Cao, Y.~Zhong, J.~Zhang, S.~Gao, and D.~Tao
\newblock  {Dearkd: Data-efficient early knowledge distillation for Vision Transformers} \newblock  In \emph{Proceedings of the IEEE/CVF Conference on Computer Vision and Pattern Recognition}, 2022, pp. 12\,052--12\,062.

\bibitem{hao2022learning}
Z.~Hao \emph{et~al.}
\newblock  {Learning efficient Vision Transformers via fine-grained manifold distillation} \newblock  \emph{Advances in Neural Information Processing Systems}, vol.~35, pp. 9164--9175, 2022.

\bibitem{bhowmick2023optimizing}
S.~Bhowmick
\newblock  {Optimizing Transformer Inference on FPGA: A Study on Hardware Acceleration using Vitis HLS} \newblock  2023.

\bibitem{yu2023boost}
C.~Yu, T.~Chen, Z.~Gan, and J.~Fan
\newblock  {Boost Vision Transformer with {GPU}-friendly sparsity and quantization} \newblock  In \emph{Proceedings of the IEEE/CVF Conference on Computer Vision and Pattern Recognition}, 2023, pp. 22\,658--22\,668.

\bibitem{reidy2023efficient}
B.~C. Reidy, M.~Mohammadi, M.~E. Elbtity, and R.~Zand
\newblock  {Efficient deployment of Transformer models on edge TPU accelerators: A real system evaluation} \newblock  In \emph{Architecture and System Support for Transformer Models (ASSYST@ ISCA 2023)}, 2023.

\bibitem{li2020ftrans}
B.~Li \emph{et~al.}
\newblock  {Ftrans: energy-efficient acceleration of Transformers using FPGA} \newblock  In \emph{Proceedings of the ACM/IEEE International Symposium on Low Power Electronics and Design}, 2020, pp. 175--180.

\bibitem{wang2022via}
T.~Wang \emph{et~al.}
\newblock  {Via: A novel vision-transformer accelerator based on FPGA} \newblock  \emph{IEEE Transactions on Computer-Aided Design of Integrated Circuits and Systems}, vol.~41, no.~11, pp. 4088--4099, 2022.

\end{thebibliography}
\bibliographystyle{IEEEtaes.bst}

\begin{IEEEbiography}[{\includegraphics[width=1in,height=2in,clip,keepaspectratio]{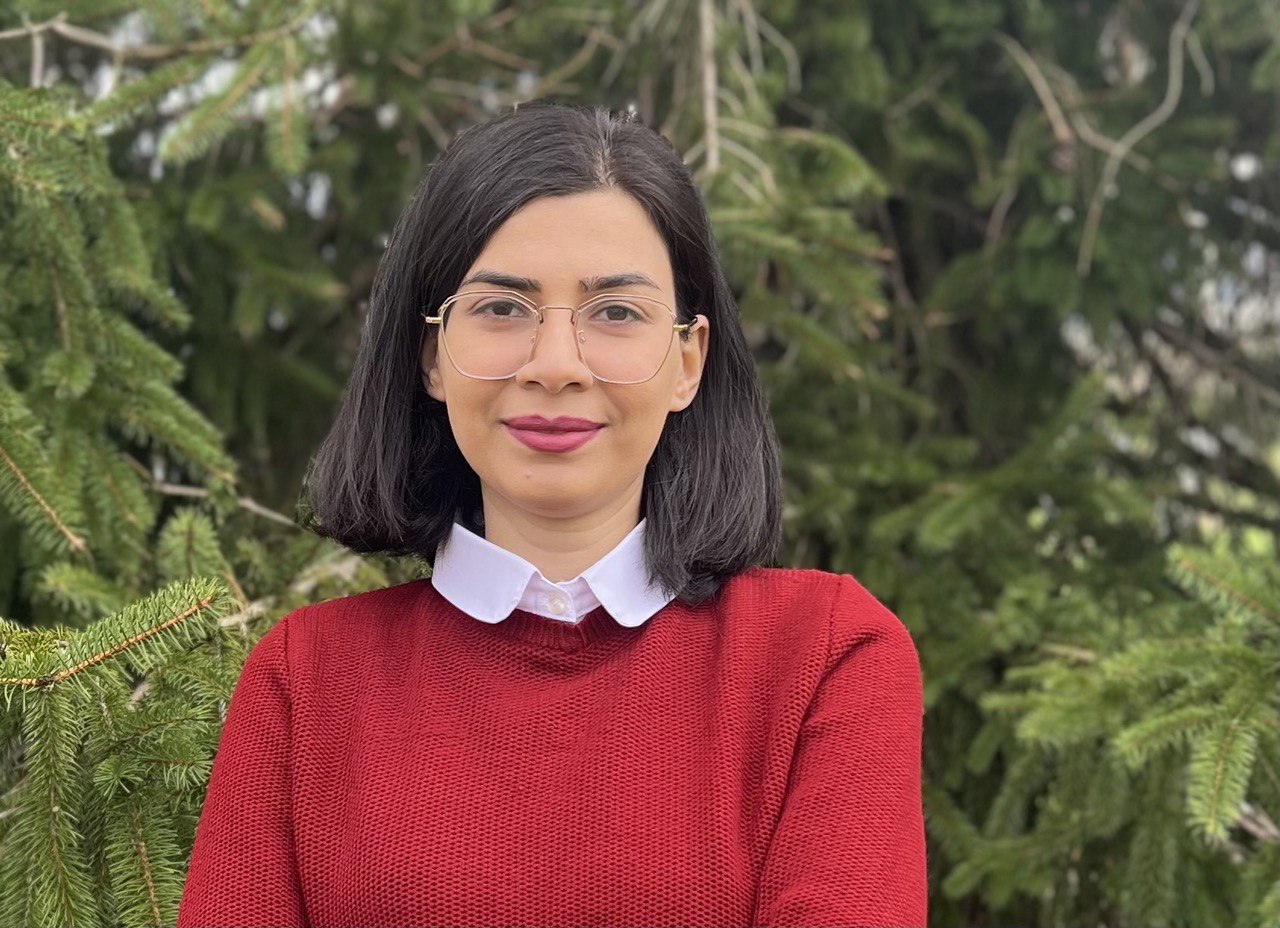}}]{Atefeh Khoshkhahtinat}{\space}received the B.Sc.  degree  in  electrical  engineering from the K. N. Toosi University of Technology, Tehran, Iran, the M.Sc. degree in electrical engineering from the Sharif University of Technology, Tehran, Iran. She is currently pursuing the Ph.D. degree with the Lane Department of Computer Science and Electrical Engineering, West Virginia University, USA. Her research interests include deep learning, machine learning, and computer vision.
\end{IEEEbiography}

\begin{IEEEbiography}[{\includegraphics[width=1in,height=1.25in,clip,keepaspectratio]{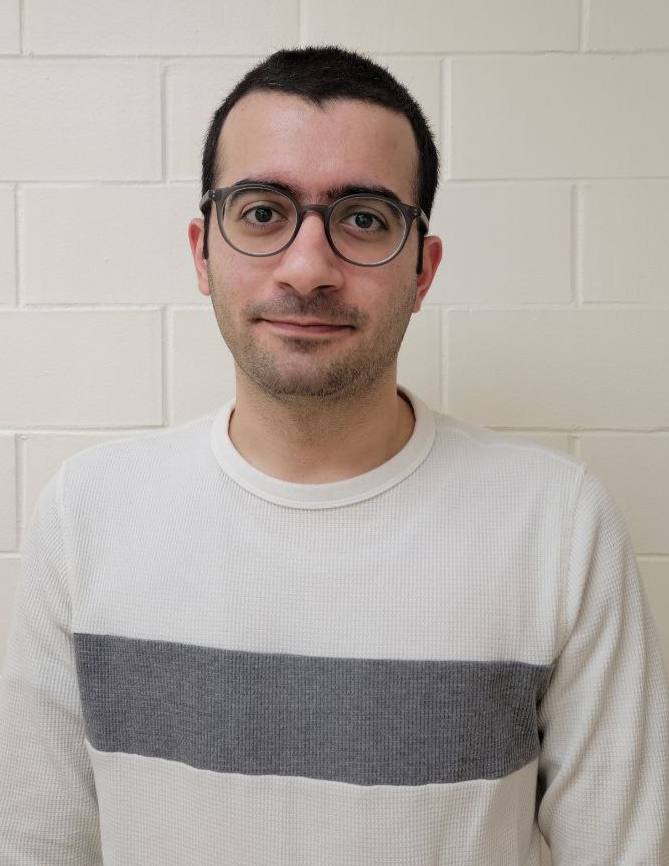}}]{Ali Zafari}{\space}received his B.Sc. (2016) and M.Sc. (2020) degrees in electrical engineering from University of Tehran and Sharif University of Technology, Tehran, Iran, respectively. He is currently pursuing a doctorate at the Lane Department of Computer Science \& Electrical Engineering at West Virginia University, where he is engaged in research related to deep learning, information theory and computer vision. In particular, He is applying this research in the field of space data compression schemes, and have worked towards improving image compression algorithms for solar missions.
\end{IEEEbiography}%

\begin{IEEEbiography}[{\includegraphics[width=1in,height=1.25in,clip,keepaspectratio]{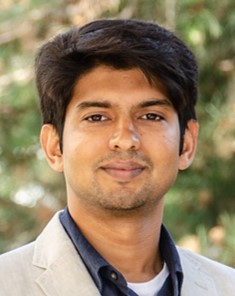}}]{Piyush M. Mehta}{\space}received his B.S. and Ph.D. degrees in Aerospace Engineering from the University of Kansas in 2009 and 2013, respectively. He is currently an assistant professor of space systems and a J. Wayne and Kathy Richards Faculty Fellow in the Department of Mechanical and Aerospace Engineering at West Virginia University. His research interests lie at the intersection of Astrodynamics, Space Science and Technology, and Data Science. His primary research area is investigating the impact of space weather on technological assets in space and on the ground. He is also interested in broader application of Machine Learning techniques to challenging problems across domains including Unsteady Aerodynamics and Data Compression. He is a member of the NASA Space Weather Council.
\end{IEEEbiography}

\begin{IEEEbiography}[{\includegraphics[width=1in,height=1.25in,clip,keepaspectratio]{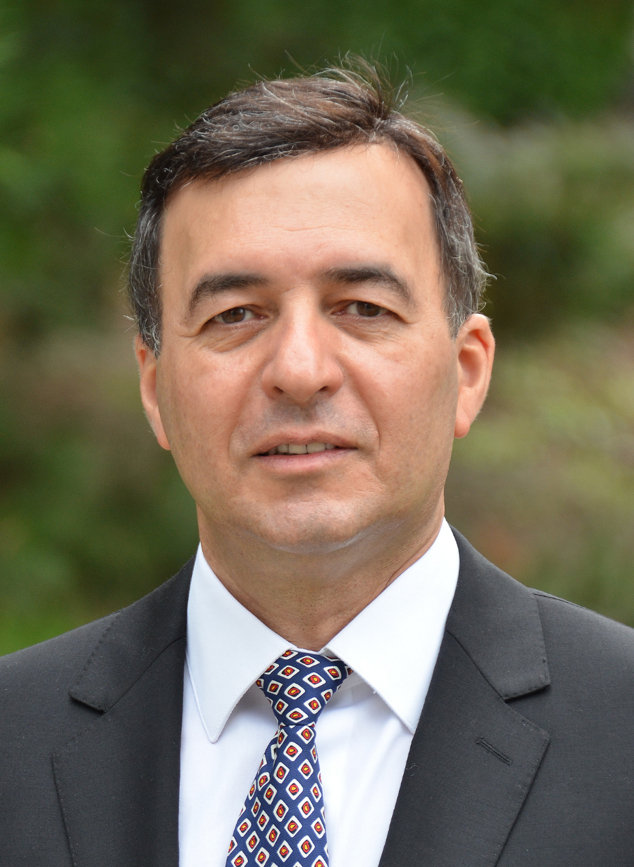}}]{Nasser M. Nasrabadi}{\space}(Fellow, IEEE) received the B.Sc. (Eng.) and Ph.D. degrees in electrical engineering from the Imperial College of Science and Technology, University of London, London, U.K., in 1980 and 1984, respectively. In 1984, he was at IBM, U.K., as a Senior Programmer. From 1985 to 1986, he was at the Philips Research Laboratory, New York, NY, USA, as a member of the Technical Staff. From 1986 to 1991, he was an Assistant Professor at the Department of Electrical Engineering, Worcester Polytechnic Institute, Worcester, MA, USA. From 1991 to 1996, he was an Associate Professor at the Department of Electrical and Computer Engineering, State University of New York at Buffalo, Buffalo, NY, USA. From 1996 to 2015, he was a Senior Research Scientist at the U.S. Army Research Laboratory. Since 2015, he has been a Professor at the Lane Department of Computer Science and Electrical Engineering. His current research interests include deep learning, computer vision, biometrics, statistical machine learning theory, and image processing. He is a fellow of the International Society for Optics and photonics (SPIE), and IEEE. He has served as an Associate Editor for the IEEE TRANSACTIONS ON IMAGE PROCESSING, the IEEE TRANSACTIONS ON CIRCUITS AND SYSTEMS FOR VIDEO TECHNOLOGY, and the IEEE TRANSACTIONS ON NEURAL NETWORKS AND LEARNING SYSTEMS.
\end{IEEEbiography}

\begin{IEEEbiography}[{\includegraphics[width=1in,height=1.25in,clip,keepaspectratio]{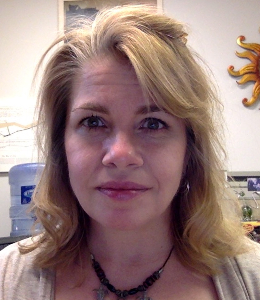}}]{Barbara J. Thompson}{\space}received a B.A. in Physics and Mathematics from the University of Pennsylvania in 1991 and a Ph.D. in Physics from the University of Minnesota in 1996.  Thompson has been a civil servant NASA Goddard Space Flight Center since 1998, and is a research scientist in the Heliophysics Science Division. Dr. Thompson  has severed on numerous space mission teams, with a  research emphasis on dynamic phenomena on the Sun and the inner Heliosphere.  Thompson serves as part of the leadership team for the Center for HelioAnalytics, a Community of Practice to envision solutions using machine learning, knowledge capture, and data analytics to expand the discovery potential for key heliophysics research topics and missions.  
\end{IEEEbiography}

\begin{IEEEbiography}[{\includegraphics[width=1in,height=1.25in,clip,keepaspectratio]{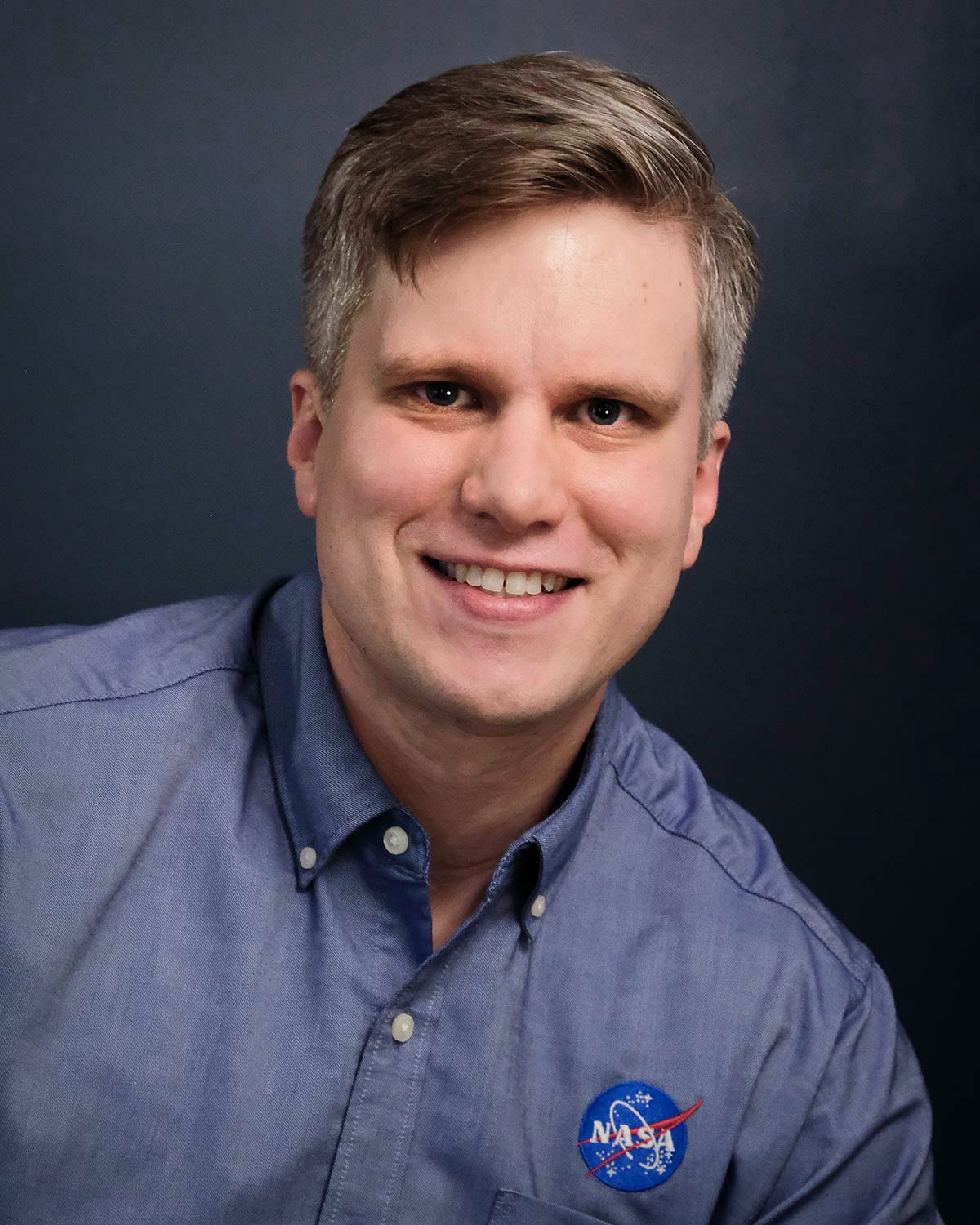}}]{Michael S. F. Kirk}{\space}was born in Salem, OR. He received a B.A. (2006) degree in physics and astronomy
from Whitman College, Walla Walla, WA, USA and M.S. (2011) and Ph.D. (2013) degrees in astronomy from New Mexico State University, Las Cruces, NM, USA.
From 2013 through 2016, he was a NASA Postdoctoral Fellow at NASA’s Goddard Space Flight Center in the Heliophysics Science Division in Greenbelt, MD, USA. From 2016 to 2021, he was a Research Scientist with The Catholic University of America in Washington, DC, USA. From 2021 to 2022, he was a Principal Scientist with Atmospheric and Space Technology Research Associates, LLC, Louisville, CO, USA. In 2022, he joined NASA’s Goddard Space Flight Center, Greenbelt, MD, USA as a Research Astrophysicist. In this position, he is the Principal Investigator of NASA's Heliophysics Education Activation Team (NASA HEAT) and a Co-investigator on the SunCET CubeSat mission, launching in 2024. In addition, he is helping to lead NASA Goddard’s Center for HelioAnalytics which seeks to integrate data science into
heliophysics to better the physics of the sun, the causes of solar variability, and its impacts on Earth.
Dr Kirk is a member of the American Astronomical Society, Solar Physics Division, where he serves as the Press Officer. He is also a member of the American Geophysical Union, Space Physics and Aeronomy section.
\end{IEEEbiography}

\begin{IEEEbiography}[{\includegraphics[width=1in,height=1.25in,clip,keepaspectratio]{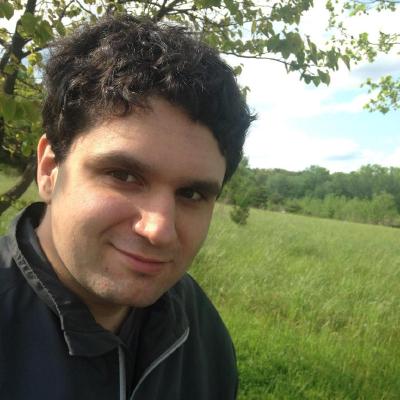}}]{Daniel E. da Silva}{\space}holds a BSc in Mathematics from University of Maryland, College Park and a MSc in Space Systems Engineering from Johns Hopkins University. He is a doctorate student in the Laboratory for Atmospheric and Space Physics at the University of Colorado, Boulder. His research interests include applied information theory and data compression, radiation belt modeling, and probabilistic space weather forecasting.
\end{IEEEbiography}

\end{document}